\documentclass[fleqn,usenatbib]{mnras}

\usepackage{newtxtext,newtxmath}

\usepackage[T1]{fontenc}

\DeclareRobustCommand{\VAN}[3]{#2}
\let\VANthebibliography\thebibliography
\def\thebibliography{\DeclareRobustCommand{\VAN}[3]{##3}\VANthebibliography}

\usepackage{mathastext}
\usepackage{subcaption}
\usepackage{graphicx}
\usepackage{caption}
\usepackage{mathtools}
\usepackage{amsmath}	
\usepackage{tikz}
\usetikzlibrary{shapes.geometric, arrows}

\tikzstyle{startstop} = [rectangle, rounded corners, 
minimum width=3cm, 
minimum height=1cm,
text centered, 
draw=black, 
fill=red!30]

\tikzstyle{io} = [trapezium, 
trapezium stretches=true, 
trapezium left angle=70, 
trapezium right angle=110, 
minimum width=3cm, 
minimum height=1cm, text centered, 
text width=4cm,
draw=black, fill=blue!30]

\tikzstyle{process} = [rectangle, 
minimum width=3cm, 
minimum height=1cm, 
text centered, 
text width=3cm, 
draw=black, 
fill=orange!30]

\tikzstyle{calculation} = [ellipse, 
minimum width=3cm, 
minimum height=1.5cm, 
text centered, 
draw=black, 
text width=3cm,
fill=violet!30]

\tikzstyle{decision} = [rectangle, 
minimum width=3cm, 
minimum height=1cm, 
text centered, 
draw=black, 
text width=3cm,
fill=green!30]

\tikzstyle{arrow} = [thick,->,>=stealth]

\makeatletter
\def\fps@figure{!ht}
\makeatother

\title[Terrestrial Planets Water Inheritance]{Water Astrochemical Inheritance of Terrestrial  Planets from Local Wet Silicates}
\title[Terrestrial Planets Water Inheritance]{Astrochemical Inheritance of Terrestrial  Planets Water from Local Wet Silicates}

\author[Boitard-Crépeau \& Pantaleone et al.]{
  Lise Boitard-Crépeau,$^{1}$\thanks{Boitard-Crépeau \& Pantaleone contributed equally to this work.}
  Stefano Pantaleone,$^{2}$
  Cecilia Ceccarelli,$^{1}$\thanks{E-mail: cecilia.ceccarelli@univ-grenoble-alpes.fr}
  Pierre Beck,$^{1}$
  Lydie Bonal$^{1}$
  \and
  Piero Ugliengo$^{2}$
\\\\
$^{1}$Univ. Grenoble Alpes, CNRS, IPAG, 38000 Grenoble, France\\
$^{2}$Dipartimento di Chimica and Nanostructured Interfaces and Surfaces (NIS) Centre, Universit\`{a} degli Studi di Torino, via P. Giuria 7, 10125, Torino, Italy
}

\date{Accepted 2026 April 21. Received 2026 April 17; in original form 2026 March 16}

\pubyear{2026}

\begin{document}
\label{firstpage}
\pagerange{\pageref{firstpage}--\pageref{lastpage}}
\maketitle

\begin{abstract}
The delivery of water to the inner Solar System rocky planets, including Earth, remains debated, as standard models assume that they formed from dry grains, inside the snowline of the protosolar nebula. However, a recent work showed that a not-negligible amount of water formed during the  prestellar phase could have been retained by pebbles and planetesimals at the Earth's orbit in enough quantities to reproduce its water content. This study was based based on quantum mechanics (QM) calculations of the binding energy (BE) of water on amorphous ice and on a kinetic approach. Here, we present new QM calculations of the BE of water frozen on the surface of silicate grains, and show that it is on average about twice larger than that on the amorphous ice. 
The contribution of this first layer of frozen water increases the dust temperature at which frozen water can be retained. This provides a local source of water not only for the Earth, but also for the inner rocky planets. The predictions from our model are in agreement with the available estimates of water content in terrestrial planets. This suggests that water delivery from the outer Solar System may not be required. 
\end{abstract}



\section{Introduction}\label{sec:intro}

The role of water in shaping Earth in the planet that we know and in the emergence of life is manifold, making water a molecule of paramount importance for the definition of extra-solar planet habitability.
Yet the origin of terrestrial water is still a debated issue.
The fundamental and unanswered question is whether the rocks that built Earth contained water, to be more precise water-equivalent compounds where H atoms are incorporated in different minerals and hydrous phases \citep[e.g.,][]{Peslier2017-SSRvEarthWater}, and if so, in what quantities.
The answer boils down to understand where the so-called snowline, i.e. where the transition solid-gaseous water takes place, was located at the time of the formation of the first pebbles, which further coalesced into planetesimals and, eventually, Earth.
If the Earth's orbit was inside the water snowline, then those first pebbles were dry and, therefore, the Earth would have been dry as well.
Traditionally, the snowline location is computed at the water condensation temperature, with values commonly found between 145 and 180 K \citep{loddersSolarSystemAbundances2003, lecarSnowLine2006, Hartmann2017-DiskWater}. 
Under this assumption, the snowline was located at more than 3 au \citep{hayashiPSN1981}, even though this value could have slightly varied, depending on the luminosity of the nascent Sun and the viscosity in its circumstellar disk \citep[e.g.,][]{Hartmann2017-DiskWater}, and, therefore, Earth would have been dry.

However, a recent work showed that the energy that binds water molecules to the ice coating the interstellar/protoplanetary disk dust grains, called binding energy (BE), has not a single value but a gaussian-like distribution that covers a relatively large range of values \citep[14 to 62 kJ/mol:][]{Tinacci2023waterBE}.
In turn, the water snowline is not a sharp transition but rather a diffuse one, caused by the gradual sublimation of water ice with increasing temperature toward smaller distances from the central (proto-)Sun.
\cite{Boitard-Crepeau2025ApJL} applied the BE distribution by \cite{Tinacci2023waterBE} to the case of the Proto-Solar Nebula (PSN) and showed that the entire terrestrial water could be inherited from the ice covering dust grains at the Earth's orbit. 
We call this scenario "astrochemical inheritance", as water ices are formed very early in the PSN history, in its prestellar core \citep[e.g.,][]{Ceccarelli2014-PP6}.
In their model, \cite{Boitard-Crepeau2025ApJL} only considered the BE due to the water molecules bound to the amorphous solid water surface, which strictly applies only to the bulk of a thick ice.
In this respect, a typical interstellar grain (with a $0.1\mu$m radius) is predicted to be coated by an ice of about 100 layers of water molecules \citep[][]{Ceccarelli2018-ices}.
However, the first ice layers enveloping the core dust silicate surfaces are likely much more strongly bound and, consequently, a larger temperature is needed to release them into the gas-phase.
While this may have a minor role in the terrestrial water astrochemical inheritance, it could have a decisive impact to the water content in Venus and, maybe, also Mercury.

In this article, we report new quantum mechanics (QM) calculations of the BE of water attached to silicate surfaces and how they impact the predicted water content of the Solar System rocky planets.
In particular, we first present the new QM calculations (Sec. \ref{sec:BE-silicate}), then describe how we model the new snowline (Sec. \ref{sec:ModelSnowline}) and finally discuss the implications for the terrestrial planets water astrochemical inheritance (Sec. \ref{sec:Discussion}). 
A final section concludes on the consequences of the present work on the presence of indigenous water in the inner planets of the Solar System (Sec. \ref{sec:conclusions}).

\section{Water binding energy on silicates}\label{sec:BE-silicate}
\subsection{General Considerations}
Interstellar grains are known to have an average radius of 0.1~$\mu$m \citep[e.g.,][]{Galliano2018-ARAA}, containing therefore several millions of atoms. Simulating such systems with QM methods is impossible with the available computing facilities. So far, the largest simulated silicate grains contain about 800 atoms \citep{zamirri2019}, whereas the largest water ice grain is made up of about 1000 atoms \citep{Germain2022}.
In our simulations, we had, therefore, to limit the size of the grains to similar values.
Specifically, we built two silicate grains, respectively obtained through annealing and nucleation of amorphous silicate nanoparticles, containing about 100 atoms and whose radius is about 5 \AA ~(as described in Sec. \ref{subsec:BEmethod}). We then cover these “naked” silicates with three monolayers (MLs) of water molecules, for a total of about 400 molecules, which corresponds to about 1400 atoms. Therefore, our final grains are constituted of about 1500 atoms (see Table \ref{tab:appendix_BE_ML} for detailed numbers).
Giving the relatively low number of adsorbing sites (less than 60 and 100 in the first and second MLs, respectively) we could not calculate a meaningful BE distribution but only the average BE for each ML.

We emphasise that, although the simulated grains are much smaller than the real interstellar ones, they catch the QM processes adequately and, consequently, the calculated average BEs are expected to be practically unaffected by the size of the assumed grain sizes. 
This is demonstrated by Fig. \ref{fig:be_convergence} that shows the average BE as a function of the ML number: the average BE converges to that of the \cite{Tinacci2023waterBE} distribution for ML$\geq2$.

Finally, to build the silicates, we considered two nanoparticles with global chemical formula of olivine (Mg$_2$SiO$_4$), but with different local compositions of olivine, pyroxene (Mg$_2$Si2O$_6$) and magnesia (MgO), as described in Sec. \ref{subsec:structure}.

\subsection{Structural models of silicates}\label{subsec:structure}

\begin{figure*}
    \centering
    \includegraphics[width=0.9\linewidth]{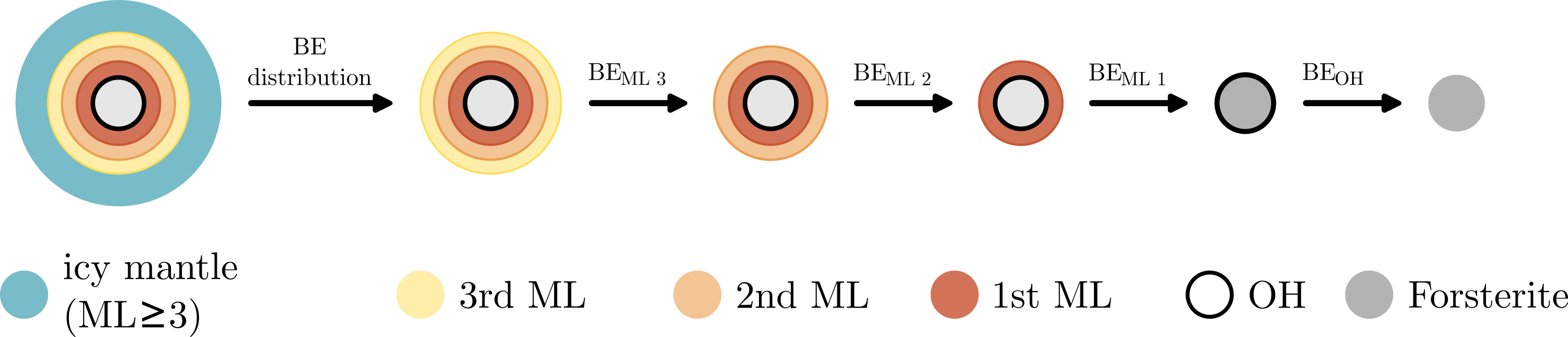}
    \caption{Schematic representation of the successive desorption of the ice coating a dust grain of silicates (central grey circle). The blue shell represents the ice bulk, formed by more than four monolayers and for which BEs are those by \protect\cite{Tinacci2023waterBE}. Yellow, light and dark orange shells represent the innermost three layers, respectively, whose BEs are calculated in this work. Spontaneously deprotonated water molecules at the forsterite surface are highlighted by the black shell.}
    \label{fig:BE_layers}
\end{figure*}

The silicate Mg$_2$SiO$_4$ core grain models used for this work were retrieved from the paper by Bromley and coworkers \citep{zamirri_nano}. 
In the present work, we selected their smallest grain, counting 14 formula units, in order to maintain computationally affordable calculations We built two different amorphous nanoparticles, which will be named throughout the paper as "annealed" and "nucleated" to identify the different approaches used to model them. 
The annealed models were obtained by high temperature molecular dynamics (MD) annealing (1800 K, 30 ps) of crystalline forsterite; the nucleated ones by monomer addition (Mg, O, SiO) over a starting seed, followed by MD simulation (1800 K, 4.5 ps). 
A key distinction between annealed and nucleated olivine nanoparticles lies in their speciation, namely the segregation into Mg$_2$Si$_2$O$_6$ and MgO. 
Although the overall chemical formula is Mg$_2$SiO$_4$ for all nanoparticles considered, three principal structural types can be distinguished at the coordination level: forsterite (Mg$_2$SiO$_4$), enstatite (Mg$_2$Si$_2$O$_6$), and magnesium oxide (MgO), coexisting depending on the equilibrium condition:
\begin{equation}
2Mg_2SiO_4 \rightleftharpoons Mg_2Si_2O_6 + 2MgO
\end{equation}
\noindent Typically, annealed nanoparticles contain forsterite as the dominant phase, since they originate from crystalline forsterite, even though minor fractions of pyroxene and MgO are also present. 
This situation arises when two silicate tetrahedra link covalently through formation of a siloxane bridge:
\begin{equation}
2SiO_4^{4-} \rightleftharpoons Si_2O_7^{6-} + O^{2-}
\end{equation}
\noindent which corresponds solely to the anionic component of the previous reaction. 
In the present study, the selected nanoparticles are composed of 14 formula units. 
The annealed particle consists of 10 Mg$_2$SiO$_4$, 2 Mg$_2$Si$_2$O$_6$, and 2 MgO units, whereas the nucleated model contains 2 Mg$_2$SiO$_4$, 5.5 Mg$_2$Si$_2$O$_6$, and 11 MgO units. 
The non-integer proportion of enstatite in the nucleated system leads to the appearance of unusual motifs such as planar SiO$_3$, disrupting the standard stoichiometry of the remaining silicate and siloxane groups. 
According to the results reported by \cite{zamirri_nano}, annealed nanoparticles are more stable than nucleated ones at all examined sizes. 
Therefore, nucleated particles are expected to be overall more reactive because they expose a greater number of ions in non-olivine environments. 
Nevertheless, the higher thermodynamic stability of annealed nanoparticles does not rule out a significant presence of nucleated forms, since under interstellar medium conditions their abundance is governed primarily by formation kinetics rather than equilibrium stability.
We then built MLs of water molecules around the silicate grains, and then probe the BE of the water layers to the grain surface (see details of water adsorption in Appendix \ref{apx:sub_workflow}).

\subsection{Computational methods}\label{subsec:BEmethod}

\subsubsection{Computational details}\label{subsubsec:BEmethod-details}

All calculations were carried out with the CP2K code \citep{Khne2020} using the r$^2$SCAN functional \citep{Furness2020} coupled with a mixed Gaussian–plane-wave scheme (plane-wave cutoff = 500~Ry) and a TZVP Gaussian basis set for the valence electrons; core electrons were described by Goedecker–Teter–Hutter (GTH) pseudopotentials. 
Van-der-Waals interactions were recovered a posteriori using Grimme's D3 dispersion \citep{Grimme2010} with Becke–Johnson (BJ) damping \citep{Grimme2011}, switching off the D3 contribution from Mg (D3(Mg=0)) according to recent benchmarks \citep{Boese2013, Pantaleone2021}. 
The SCF convergence threshold was set to $\Delta E = 10^{-8}$~hartree.
A two-step geometry-optimization protocol was employed. 
In the first round, all forsterite atoms were fixed and only the H$_2$O molecule(s) were relaxed using default convergence parameters. 
In the second round, all atoms were free to relax and the optimization thresholds were tightened: Maximum and RMS (Root Mean Square) gradients equal to $1\times10^{-5}$ and $5\times10^{-6}$~hartree/bohr, respectively, and Maximum and RMS displacements equal to $1\times10^{-5}$ and $5\times10^{-6}$~bohr, respectively.

\subsubsection{Binding energy calculation}\label{subsubsec:BEmethod-calcul}

We calculated the BE of layers composed by many water molecules, where the average contribution of all waters is mediated within the whole layer, in the following way:
\begin{equation}
    \centering
    BE = \frac{(nE_{H_2O} + E_{H_2O-forst}) - E_{cplx}}{n} + \Delta ZPE^{cry}
    \label{eq:be}
\end{equation}
\noindent where $E_{H_2O}$ is the energy of the isolated $H_2O$ molecule and $E_{H_2O-forst}$ is the energy of the hydrated silicate nanoparticle, which changes with the coverage regime dictated by $E_{cplx}$.
From a physical point of view $E_{H_2O-forst}$ is the reference to calculate the energetic cost to desorb the n$^{esimal}$ water layer. 
It is not univocal for all the layers, but it depends on the hydration level of $E_{cplx}$ as depicted in Fig.~\ref{fig:BE_layers}. 
As an example, to calculate the BE of the third layer, the energy of the complex $E_{cplx}$ refers to the hydrated silicate nanoparticle coated by 3 layers of water molecules, while the reference $E_{H_2O-forst}$ refers to the hydrated silicate nanoparticle coated by 2 layers of water molecules.
$\Delta ZPE^{cry}$ is the shift applied to all BEs to include the zero-point energy (ZPE) correction, explicitly calculated on the crystalline proton ordered P-ice, and benchmarked versus the hexagonal ice (see details in Appendixes \ref{apx:sub_workflow} and \ref{apx:sub_comp_exp}).

\subsubsection{Rate constant and pre-exponential factor}\label{sec:prefactor}

The BE is a crucial parameter that determines, at a given dust temperature, whether a water molecule remains frozen on the grain, or it sublimates into the gas-phase. 
This is described by the thermal desorption rate $k_{des}$, defined by:
\begin{equation}
    k_{des} ~=~ \nu ~\exp\left(-\frac{BE}{T}\right) \label{eq:desorptionrate}
\end{equation}
\noindent where $\nu$ is the pre-exponential factor, or prefactor, (in s$^{-1}$), $BE$ is the water binding energy (in K), and $T$ the temperature of the grain surface (in K), assumed to be equal to that of the gas. 
The prefactor is determined for each BE at the peak temperature of the desorption rate, where $\nu$ is expressed using the Tait equation \citep{tait2005} :
\begin{equation}
    \nu^{Tait} = \frac{k_B~T}{h}~\left[\frac{2~\pi~m_{H_2O}k_B~T}{h^2}A\right]~\left[\frac{\sqrt{\pi}}{\sigma~h^3}(8~\pi^2~k_B~T)^{\frac{3}{2}}~\sqrt{I_x~I_y~I_z}\right],\label{eq:prefactor}
\end{equation} 
\noindent where $k_B$ and $h$ are the Boltzmann and Planck constants and $A$ is the surface area per molecule (usually $10^{-19}~\text{m}^2$); $m_{H_2O}$, $\sigma$ and $I_i$ are the mass, symmetry factor and principal moments of inertia of the water molecule, equal to 18~a.m.u., 2 and (1.83, 1.21, 0.62)~a.m.u.~\AA$^2$, respectively. 
In the case of the BE distribution of water on ice \citep{Tinacci2023waterBE}, this equation was corrected to take into account the coupling between the adsorbed water molecule and the icy surface. 
Due to the "soft" nature of the water ice surface, vibrational modes can reduce the prefactor value by up to a factor 10 at high temperatures \citep{Tinacci2023waterBE, pantaleonePreFactor2025}. 
In the case of silicate grains, however, the structure of the surface is not expected to change upon water adsorption and, therefore, we do not need to include this correction. 
According to the study of \cite{pantaleonePreFactor2025} on prefactors, Eq. (\ref{eq:prefactor}) represents the upper bound for the value of our prefactors, but we still investigated the influence of a factor 3 times smaller on the desorption temperature.

\subsection{Binding energy Results}\label{subsec:BEresults}

\subsubsection{$H_2O$ single adsorption}\label{subsubsec:BEresults-single}

The adsorption of single water molecules was modelled on the bare silicate nanoparticles in order to study the eventual spontaneous deprotonations, which only occurs on some reactive MgO sites exposed at the surface of the grain.
Therefore, we generated the corresponding hydroxylated nanoparticles, which are used as starting seed to model the first layer of molecularly adsorbed $H_2O$. 
All details about the sampling are available in Appendix~\ref{apx:sub_h2o_ads}. 
On the annealed nanoparticle only one reactive MgO sites was identified able to spontaneously deprotonate $H_2O$ with BE = 262.7~kJ~mol$^{-1}$, while on the nucleated one up to 3 molecules deprotonate, with an averaged BE = 179.4~kJ~mol$^{-1}$.

\subsubsection{$H_2O$ monolayer to multilayer}\label{subsubsec:subsubsec:BEresults-multi}

The water MLs were built up with the SOLVATOR tool in the ORCA program: for the first ML 100 H$_2$O molecules were set in the docking algorithm in order to ensure a full coverage of all the possible adsorption sites, both Lewis acidic, i.e. undercoordinated Mg$^{2+}$ cations, and Brönsted basic sites, i.e. O atoms of silicate groups, respectively acceptors of dative bonds (from H$_2$O lone pairs) and H-bonds. 
After geometry optimization, a cleaning procedure was applied to remove H$_2$O molecules in excess. 
As the actual definition of ML can be somehow arbitrary, we defined as ML all H$_2$O molecules directly connected to silicate atoms; chemisorbed H$_2$O, i.e. spontaneously deprotonated (during the first run of adsorption), are considered as belonging to silicate. 
See details in Appendix~\ref{apx:sub_h2o_multi}.

\begin{figure}
    \centering
    \includegraphics[width=\columnwidth]{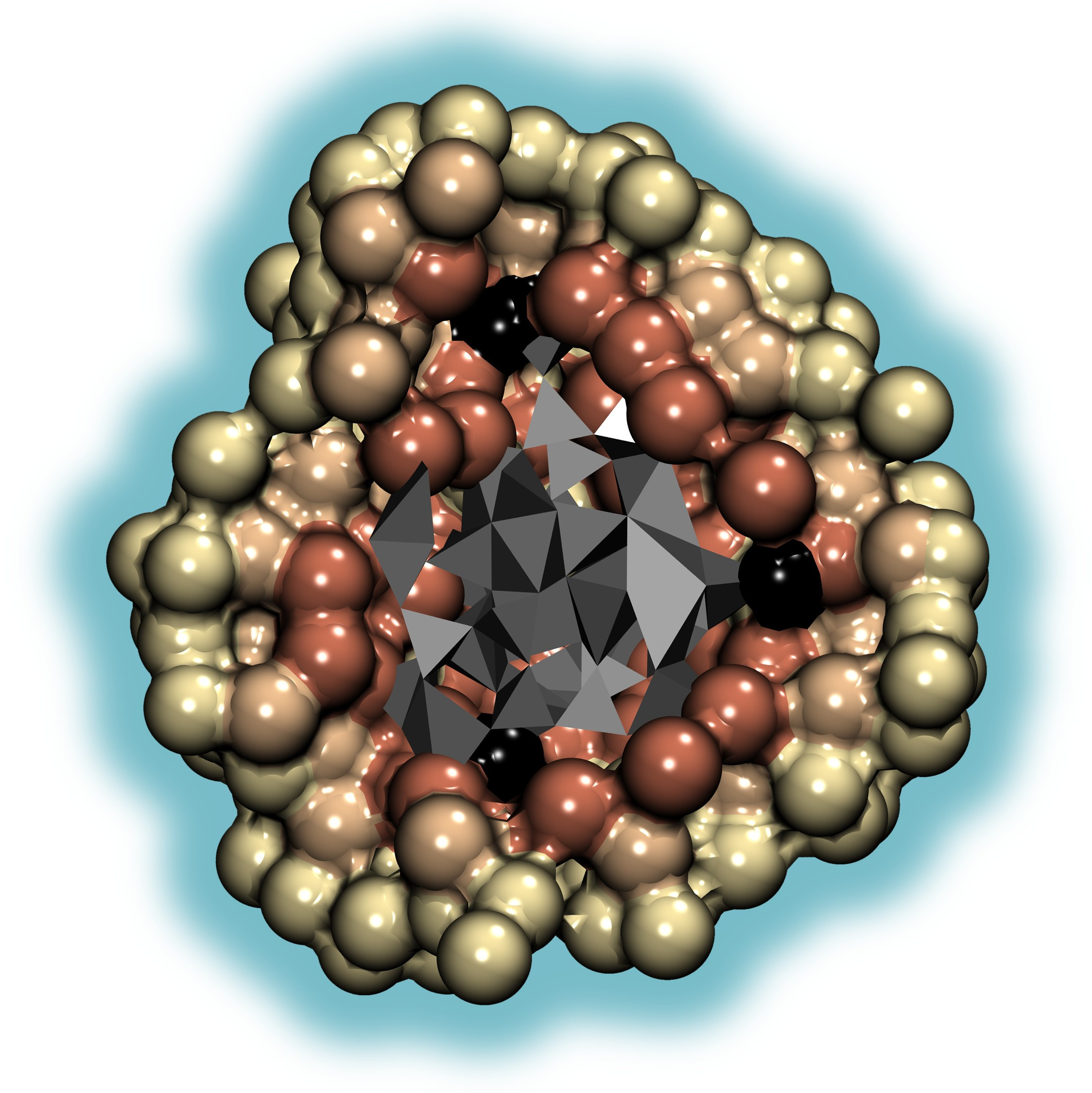}
    \caption{Maximum level of hydration of the nucleated nanoparticle. 
    Each water layer is highlighted with a different colour (only oxygen atoms are shown): spontaneously deprotonated waters in black, first ML in dark orange, second ML in light orange, third layer in yellow, $>$ 3 ML in blue (waters not explicitly modelled). 
    The silicate core atoms are represented as grey polyhedra.}
    \label{fig:h2o_multilayer}
\end{figure}

\begin{figure}
    \centering
    \includegraphics[width=\columnwidth]{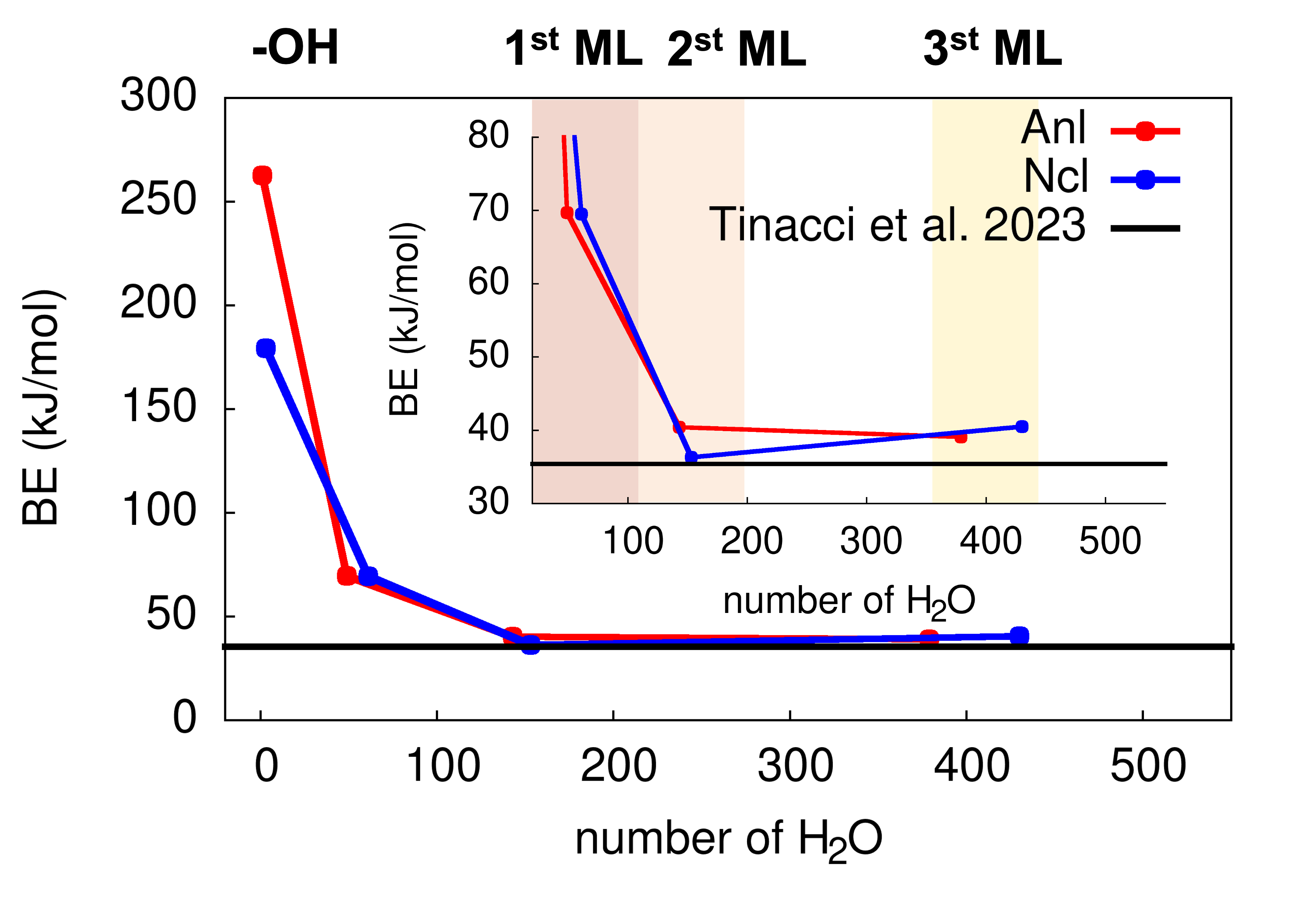}
    \caption{Convergence of the averaged BE of water versus the number of H$_2$O molecules adsorbed on annealed (Anl) and nucleated (Ncl) nanoparticles. 
    The inset chart shows the 30-80~kJ~mol$^{-1}$ range, i.e. excluding the BE of chemisorbed (i.e. dissociated) waters. 
    The value from \protect\cite{Tinacci2023waterBE} ($\mu$ = 35.4 kJ mol$^{-1}$) represents our reference BE of ice bulk.}
    \label{fig:be_convergence}
\end{figure}

We iterate the addition of further water MLs up to three units, as shown in Fig.~\ref{fig:h2o_multilayer}. 
Figure ~\ref{fig:be_convergence} reports the BE value calculated in each of the water ML shown in Fig.~\ref{fig:h2o_multilayer}. 
The most interesting and important trend which can be identified is that the higher the level of hydration, i.e. the higher the number of water layers, the lower the binding energy is. 
In other words, the closer the water molecules to the silicate nanoparticle, the higher their binding energy is. 
This behaviour is expected because the farthest water molecules with respect to silicate will interact only with each other, thus converging to the BE of the ice bulk, being it either amorphous or crystalline. 
In the (first) ML case, the range of BEs is around 62--70~kJ~mol$^{-1}$, which is close to the upper limit of the BE distribution of the single H$_2$O adsorptions (without considering chemisorption cases). 
In the second ML, BE drops down to 36--40~kJ~mol$^{-1}$; in the third ML, it does not seem to decrease, but rather to narrow the range to 39-40~kJ~mol$^{-1}$, converging to the average value from the BE gaussian distribution of \cite{Tinacci2023waterBE} (35.4~kJ~mol$^{-1}$). 

\section{Modelling of water inheritance in terrestrial planets}\label{sec:ModelSnowline}

\subsection{Model description}\label{subsec:astro-model}

From a multitude of galactic studies, we know that most of interstellar water is formed during the prestellar core phase on the surface of dust grains \citep[e.g.][]{Dulieu2010, Ceccarelli2014-PP6}.
By the end of the prestellar phase, grains are covered by thick \citep[about 100 ML, depending on the physical conditions and grain sizes: e.g.,][]{Taquet2012-GRAINOBLE, Ceccarelli2018-ices} icy mantles. 
With the evolution of the prestellar core into a protostar and protoplanetary disk, the temperature of the dust evolves and, when the temperature is larger than the desorption temperature, a fraction of ice is sublimated and only the rest remains frozen. 
Therefore, how much water remains frozen on the grains depends on the balance between the rates of water desorption and adsorption, which, in turn, depend on the temperature and density profiles across the PSN disk.


In the following, we adopt the model described in \cite{Boitard-Crepeau2025ApJL}, modified to take into account the different layers of ice coating dust grains. In this previous work, the desorption of the bulk ice was modelled using the distribution of BEs computed by \cite{Tinacci2023waterBE}, which refers to water molecules adsorbed on a grain of 200 water molecules: in other words, we assume that the \cite{Tinacci2023waterBE} BE distribution describes the BE of ML$\geq3$, which we consider to be the ice bulk.
With our new calculations, we are able to explicitly include the contribution of the first two MLs in direct contact with the silicate surface, whose first ML BEs are sensitively larger than those of the ice bulk.
Since these first two MLs contain water molecules that remain frozen at larger dust temperatures  than those of the ice bulk MLs, it is important to have the correct amount of water in those two MLs. They depend on the surface of the dust grains and, therefore, it is important here to introduce the grain radius distribution of the interstellar dust grains, which we adopt to be the canonical Mathis-Rumpl-Nordsieck one (see Sec. \ref{subsec:desorption}). Note that we also assume that the silicate mass dust-to-gas ratio remains constant during the interstellar grain coagulation into pebbles, when water ice is trapped inside.

\subsubsection{Physical structure of the Proto Solar Nebula}

The quantity of water frozen onto the dust grain depends on the gas density and temperature across the PSN midplane disk. 
As in our previous study, we used the classical Minimum Mass Solar Nebula power law \citep{weidenschilling_1977} for the gas column density and the prescription for a passively heated disk from \cite{chiang_goldreich_1997} for the temperature profile:
\begin{gather} \label{eq:PSNdens}
    \Sigma(r) = \Sigma_{1\text{au}} ~\left(\frac{r}{1 \ \text{au}}\right)^{-3/2} \ \text{g/cm$^2$} \\
    T(r) = T_{1\text{au}} \left(\frac{r}{1 \ \text{au}}\right)^{-3/7} \ \text{K} \label{eq:PSN_T},
\end{gather}
where $r$ is the distance in au from the Sun, $\Sigma_{1\text{au}}$ is the gas column density at 1 au distance, assumed to be 1700~g/cm$^2$ \citep{hayashiPSN1981}, and $T_{1au}$ is the temperature at 1~au in K. 
$T_{1au}$ changed through time with the change in solar luminosity and is kept as the only free parameter of the PSN physical structure.
This represents the grain temperature at which most of the interstellar grains coagulated into pebbles. The trapped water then is mostly conserved during the subsequent PSN evolution.

\subsubsection{Water desorption and adsorption rates} \label{subsec:desorption}

The thermal desorption rate of frozen water $k_{des}$ was introduced in Sec. \ref{sec:BE-silicate} and it is given by Eq. (\ref{eq:desorptionrate}). 
Here we use BEs of 260, 70 and 40~kJ/mol for the chemisorbed, first and second ML of water respectively. 
Desorption of the other layers of the ice is described by the BE distribution from \cite{Tinacci2023waterBE}, ranging from 14.2 to 61.6~kJ/mol.
The BEs relevant to this study are reported in Table \ref{tab:BE_mu} along with their calculated prefactors.

\begin{table}
    \caption{Adopted BE and prefactor $\nu$ for the thermal desorption rate of the different monolayers of water ice.}
    \label{tab:BE_mu}
    \centering
    \begin{tabular}{lccl}
        \hline
         Layer  & BE in kJ/mol (K)  & $\nu$ in s$^{-1}$ & Reference\\
         \hline
         -OH$^a$  &  260 (31270)      &  $8.4\times 10^{17}$ & This work\\
         1st ML &  70 (8420) &  $1.3\times 10^{16}$ & This work\\
         2nd ML &  40 (4810) &  $2.2\times 10^{15}$ & This work\\
         ML $n\geq3$ & $14.2\leq BE \leq 61.6$ & $10^{12}\leq \nu\leq 10^{14}$& $^b$\cite{Tinacci2023waterBE}\\
         \hline
    \end{tabular}
\textit{Notes:} $^a$Deprotonated water molecules. 
\end{table}

While the bulk of the water ice desorption does not depend on the grain sizes, the first layers do, because the smaller the grain the lower the number of water molecules belonging to the first layers. 
Therefore, we computed the adsorption on the grain surfaces as a function of the grain radius $a_{grain}$. 
It holds:
\begin{equation} \label{eq:k_ads}
    k_{ads} = S_{grain} ~\sigma_{grain} ~n_{grain} ~v_{th},
\end{equation}
where $S_{grain}$ is the sticking coefficient, here taken equal to 1, $\sigma_{grain}$ is the geometrical cross section of the grains equal to $\pi~a_{grain}^2$, $n_{grain}$ is the grain number density, and $v_{th}$ is the water thermal velocity equal to $\sqrt{2~k_B ~T ~/~ m_{H_2O}}$. 
In the present model, we adopt the Mathis-Rumpl-Nordsieck (MRN) grain size distribution \citep{MRN_grain_size1977}, widely used to describe the dust in the diffuse interstellar medium: 
\begin{equation}\label{eq:MNR}
    dn_{grain}(a_{grain}) = A_{sil} ~ n_{H}~ a_{grain}^{-3.5}da_{grain}
\end{equation}
where $n_H$ is the number densities (in cm$^{-3}$) of hydrogen nuclei respectively, $A_{sil}$ is a normalisation factor, equals to $7.8\times10^{—26}$ cm$^{2.5}/$H-atom in the interstellar medium (ISM) \citep{draine1984} and $a_{grain}$ is the grain radius (in cm), with $50\text{~\r{A}} < a_{grain}<0.25~\mu m$ in the ISM. We investigated the influence of grain size by multiplying these bounds by a factor 10. 
Since we assume that the silicate mass dust-to-grain ratio remains constant during grain growth, $A_{sil}$ was adjusted to $2.5\times10^{—26}$~cm$^{2.5}/$H-atom in this case.

\subsubsection{Inherited frozen water}\label{subsec:inheritance_model}
The amount of water that remains frozen onto the grain mantles at a given temperature is calculated from the equilibrium between thermal desorption of frozen water and adsorption of gaseous water onto the grains:
\begin{equation}
    \begin{cases}
    k_{ads} ~n_{H_2O, gas} ~=~ k_{des} ~n_{H_20,ice}\\
    n_{H_2O, gas} ~+~ n_{H_2O,ice} ~=~ n_{H_2O,tot} \label{eq:chem}
\end{cases}
\end{equation}
where $n_{H_2O, gas}$, $n_{H_2O,ice}$ and $n_{H_2O,tot}$ are the number densities (in cm$^{-3}$) of gaseous water, frozen water and of the total amount of water respectively. 
We solve these coupled equations for two different cases: when the desorption is from the first two ML of ice and when it is from the bulk, where we use the BE calculated in Sec. \ref{sec:BE-silicate} and by \cite{Tinacci2023waterBE}, respectively.
Then, at each temperature, which corresponds at a distance $r$ via Eq. \ref{eq:PSN_T}, we add up the two contributions.

In practice, for the first ML, the total amount of water $n_{H_2O,tot}$ is given by the number of sites that water molecules can occupy on the grain surface, which depends on the grain radius distribution (Eq. \ref{eq:MNR}), as follows: 
\begin{equation}\label{eq:n_h2o-ML}
    n_{H_2O,tot} = n_{H_2O,ML1} = \sum_{a_{grain}}\frac{4\pi~a_{grain}^2}{\pi~r_{H_2O}^2}~n_{grain}(a_{grain}),
\end{equation}
where we adopted the value $r_{H_2O}$ equal to 1.35~\r{A}, based on the study by \citet{Germain2022-iceGrain}, which shows that the average distance between two oxygen atoms of amorphous iced water is 2.7 ~\r{A}. 
Similarly, $n_{H_2O,ML2}$ in the second ML is obtained increasing $a_{grain}$ by the thickness of the first ML (i.e. the size of a water molecule).

For the remaining layers, since the bulk of the ice is described by a distribution of BE, Eq. (\ref{eq:chem}) is solved for each BE and $n_{H_2O, tot}$ is weighted by the associated fraction of ice \citep[for more details, see][]{Boitard-Crepeau2025ApJL}. The total amount of water $n_{H_2O,tot}$ in the bulk of the ice is given by the total frozen water, given by $A_{H_2O}~n_{H}$, minus the frozen water in the two first MLs:
\begin{equation}\label{eq:n_h2o-bulk}
     n_{H_2O, tot} = n_{H_2O, ML\geq3} ~=~ A_{H_2O} ~n_{H} - n_{H_2O,ML1} - n_{H_2O,ML2}
\end{equation}
where $A_{H_2O}$ is the water abundance and $n_{H}$ the number density of H nuclei. 

In the Solar System, the comet 67P/Churymov-Gerasimenko is considered one of the most pristine remnant of the earliest stages. 
We therefore chose to set $A_{H_2O}$ to match the water to rock ratio measured in this comet by \cite{fulle2019}, namely 25~wt\% of ice, equivalent to $A_{H_2O}=1\times10^{-4}$. 
Note that a wide range of values of the refractory-to-ice mass ratio are found in the literature for this comet \cite[see, e.g., the recent review by][]{marschall2025}. 
That said, this parameter only sets the amount of water contained in the bulk of the ice (Eq. \ref{eq:n_h2o-bulk}), which is of importance mainly for the cold grains in the outer PSN disk. 
This work focuses on the inner Solar System planets, in the hotter region dominated by the desorption of the first MLs (see next Section), so that the uncertainty on $A_{H_2O}$ does not sensitively impact the results of this work. 
This is briefly discussed at the end of Sec. \ref{subsec:astro-results}, and shown in Fig. \ref{fig:terrestrial_water_appendix}. 

In the following, we will describe the desorption processes in terms of water-equivalent content, expressed in weight \% (wt.\%) with respect to the total mass of silicate and ice.

\subsection{Results}\label{subsec:astro-results}

\begin{figure}
    \centering
    \includegraphics[width = \linewidth]{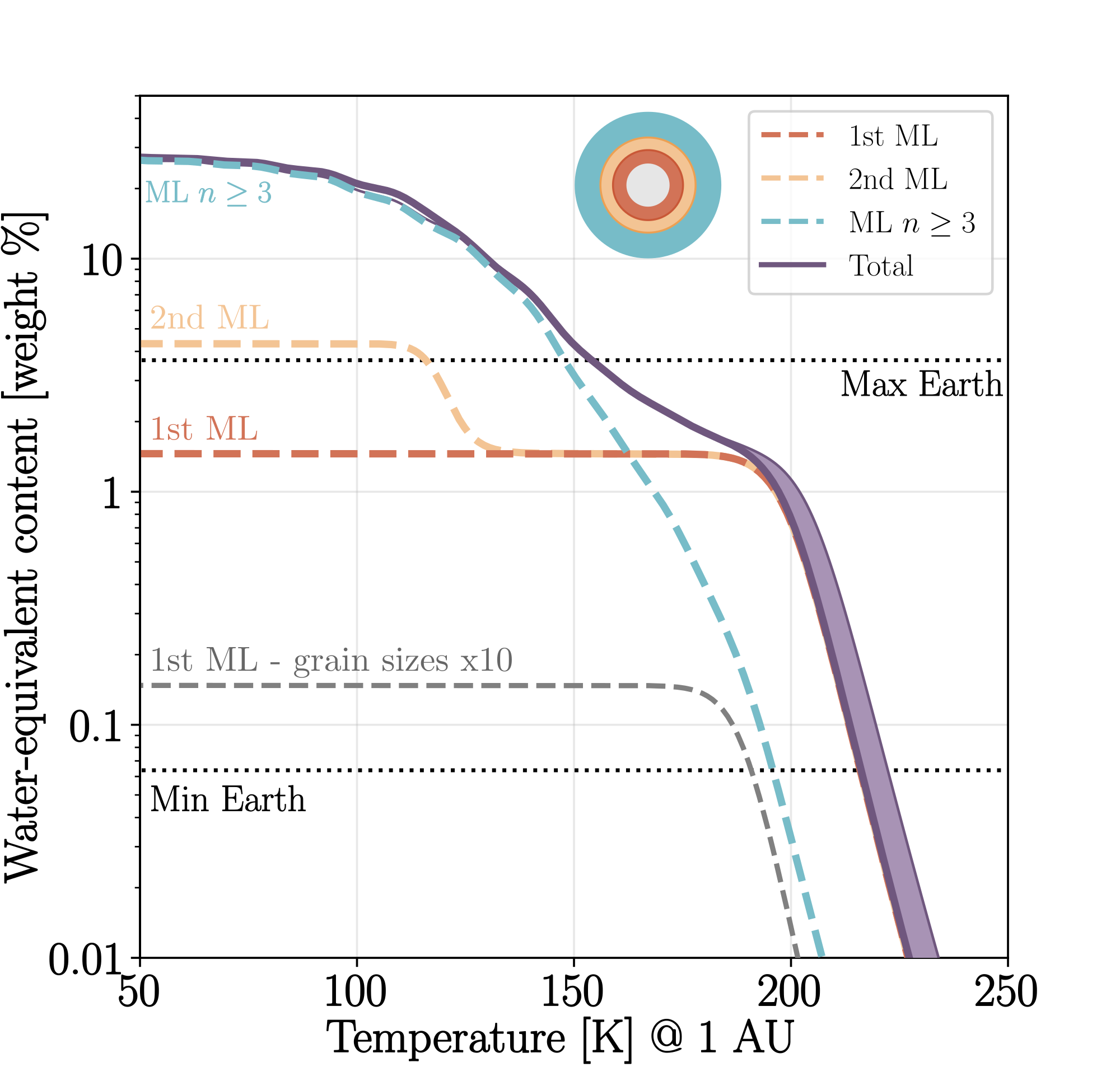}
    \caption{Ice to grain mass ratio as a function of temperature $T_{1au}$ at the Earth's orbit (1~au).
    The purple curve shows the total amount of frozen water; 
    the light and dark orange dashed lines show the frozen water of the second and first ML, respectively;
    the blue dashed line corresponds to the frozen water in the bulk layers ($\geq 3$ ML).
    The filled purple area shows to uncertainty on the prefactor value, if it is decreased by less than a factor 3 (see text). 
    The dashed light grey line shows the amount of water in the first ML when the grain sizes are increased by a factor 10 (see text).
    The minimum (0.0638 wt.\%) and maximum (3.6 wt.\%) estimates of the Earth's water content are represented as dotted black horizontal lines \citep{Peslier2017-SSRvEarthWater}. }
    \label{fig:snowlines}
\end{figure}

Figure \ref{fig:snowlines} shows the predicted water-equivalent content (in weight \% with respect to the total mass of silicate and ice) at the Earth's orbit as a function of the assumed temperature $T_{1au}$ (Eq. \ref{eq:PSN_T}). 
The figure reports the total amount of iced water (purple curve): it is composed by the iced water in the bulk (namely in the $\geq 3$ ML: light blue dashed curve) plus that in the first two ML (light and bright orange dashed curves).

The low temperature end of Fig. \ref{fig:snowlines} reflects the initial conditions of the icy grains in the ISM, before the onset of planetary formation. 
When all of the water is frozen on the grains, $\sim$1.5~wt.\% of water molecules are found in the 1st ML, and $\sim$4~wt.\% are found in the 2nd ML. 
As temperatures increase in the PSN, the bulk of the ice (light blue curve) progressively desorbs. 
This behaviour is explained in \cite{Boitard-Crepeau2025ApJL}, and referred as a \textit{diffuse snowline}. 
The 2nd ML desorbs around 120~K, but has a BE similar to the peak of the BE distribution by \cite{Tinacci2023waterBE}: its contribution to the total snowline is therefore negligible. 
On the other hand, the BE of the 1st ML (70~kJ/mol: this work) is larger than the highest BE of the distribution \citep[62~kJ/mol:][]{Tinacci2023waterBE}. 
The onset of the desorption of the 1st ML (dark orange curve) occurs around 200 K, when the bulk of the ice has been mostly desorbed. 
When compared to the estimates by \cite{Boitard-Crepeau2025ApJL} of the Earth’s water content, the contribution of the 1st ML allows water retention at higher temperatures.


To understand the robustness of the predictions, we varied the prefactor $\nu$ (Eq. \ref{eq:prefactor}), decreasing it by a factor 3, following the discussion in Sec. \ref{sec:prefactor}. 
As a result, the total amount of iced water is slightly shifted towards higher temperatures (purple shadowed curve).

The mass fraction of water contained in the 1st and 2nd MLs do not depend on the chosen water abundance, but on the sizes of the grain (Eq. \ref{eq:n_h2o-ML}). 
Therefore, we verified the impact of increasing the sizes of the grains, as it can be expected to occur on the PSN midplane, by multiplying by a factor 10 the minimum and maximum grain radius $a_{grain}$ in the MRN distribution (Eq. \ref{eq:MNR}).
In this case, the impact on the frozen water of the first ML (grey dashed curve) is large, bringing it to about 10 times lower than with the standard MRN distribution.
This reflects both the decrease in number density of larger grains (Eq. \ref{eq:MNR}) and their lower surface area to volume ratio, which reduces the relative mass contribution of water MLs compared to smaller grains. 
Since we assume that the ISM silicate mass dust-to-gas ratio remains constant, this leads to an overall decrease in the mass fraction of water in the first layer with increasing grain size.

A final additional source of water is represented by the water chemisorbed on the silicates.
As shown in Sec. \ref{subsubsec:BEresults-single}, water is deprotonated only on reactive MgO sites on silicate surfaces, resulting in very high BE which correspond to the formation of a chemical covalent bond with the surface.
For instance, deprotonated water molecules on the annealed silicate, with a BE of 262.7 kJ~mol$^{-1}$, can only be removed when the grains are heated above temperatures of order $\sim$700 K.
Unfortunately, quantifying the number of deprotonated water molecules on the grain surfaces is not trivial, as in our calculations we considered a small grain with $a_{grain} \sim$ 5~\AA, which contains a single reactive MgO site in the case of the annealed silicate. 
For a first-order, rough estimate of the number of deprotonated water molecules in a generic grain with radius $a_{grain}$, we can assume one deprotonated H$_2$O molecule per an area corresponding to a radius of 5~\AA~ and, hence, obtain the total deprotonated waters as $\left( \frac{5 \text{~\r{A}}}{a_{grain}} \right)^2$.
Using an MRN grain size distribution between 50~\AA{} and 0.25 $\mu$m, we find that the mass fraction of chemisorbed water on amorphous ice is 0.027 wt.\%, slightly below the minimum water content of the Earth (0.064~wt.\%; see next section).
The contribution of the chemisorbed water may thus not be negligible. 
However, more accurate estimates require new very computationally-expensive calculations on larger grains, which is beyond the scope of this article. 

\section{Discussion}\label{sec:Discussion}

The new calculations of the BE of the first ML of iced water coating dust grains demonstrate that these layers would remain frozen at relatively large temperatures.
For example, in the previous section we showed that about 1.5~wt.\% of frozen water stay attached to the silicate surface up to a temperature of $\sim 200$ K (Fig. \ref{fig:snowlines}). 
This opens the possibility that not only Earth and Mars but also the other inner rocky planets of the Solar System, Venus and Mercury, may have inherited a not negligible amount of astrochemical water.

\begin{figure*}
    \centering
    \includegraphics[width=0.8\linewidth]{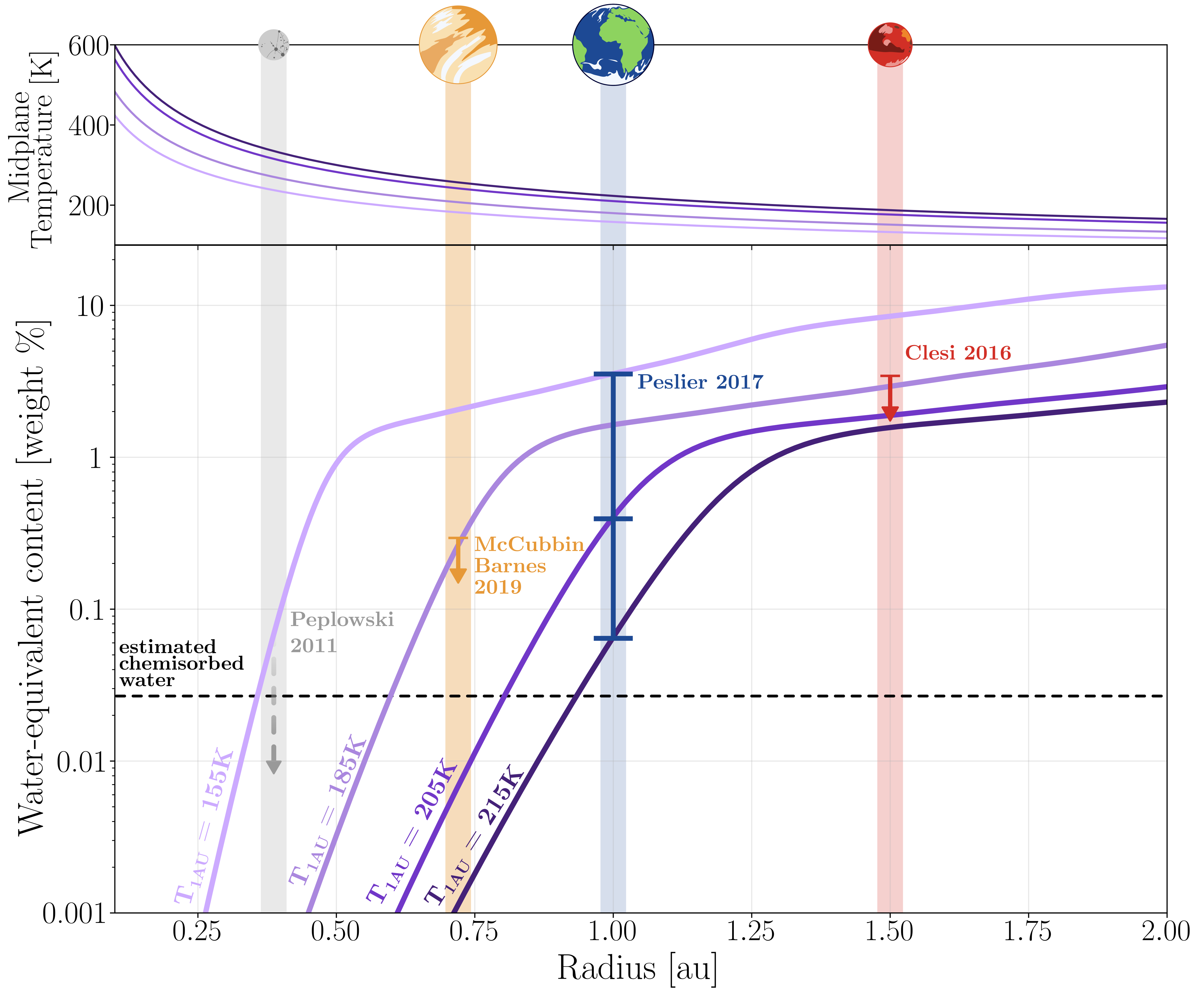}
    \caption{Model prediction for the water budget of the 4 terrestrial planets : Mercury in grey, Venus in  yellow, the Earth in blue and Mars in red. \textit{Upper Panel:} Midplane temperature (K) across the PSN following Eq. (\ref{eq:PSN_T}) with $T_{1~AU}=$ 160, 190, 215 and 225~K from the lowest (light green) to the highest (dark green) curve. \textit{Lower Panel:} Mass fraction of water remaining frozen on the grains as a function of the distance from the Sun, according to our desorption model. Each curve show a different snowline depending on the temperature profile set by $T_{1~AU}$, with colours matching the profiles of the upper panel. 
    Upper limits for the planets water contents are represented as downward arrows, using values from references cited in Section \ref{sec:water_planets}. 
    The Earth's water estimate is more constrained, and its range is represented as a vertical bar. 
    The curves with $T_{1~AU}=$160, 215 and 225~K meet the upper, median and lower values of the Earth's water estimate respectively. The dashed horizontal line shows the estimate of deprotonated water, chemisorbed to the forsterite surface (not taken into account in our diffuse snowlines).}
    \label{fig:terrestrial_water}
\end{figure*}

\subsection{Water content of terrestrial planets}\label{sec:water_planets}


The water content of the terrestrial planets, is not well constrained, especially when it comes to estimate the amount of water initially accreted by the planets. 
In the following, we review the estimates of their water-equivalent content.
More comprehensive reviews on the water content of terrestrial planets are provided by \cite{PeslierdeSanctis2022} and \cite{greenwood2018}.
Since the water-equivalent content in the terrestrial planets is highly model-dependent, we considered the upper limits of the estimated values for comparison with the predictions from our model.

\paragraph*{Earth:}
Earth's water from the hydrosphere (atmosphere and oceans) only represents a small fraction of the H$_2$O budget compared to the water-equivalent in the inner Earth. A recent estimate by \cite{Peslier2017-SSRvEarthWater} provides a total terrestrial water-content between 0.064 and 3.7 wt.\%, with a median at 0.39~wt.\%. The main uncertainty lies in how much water equivalent is contained in the core of the Earth. A recent experimental study \citep{huang2026} suggested that up to an equivalent of 45 Earth's ocean mass could be contained in the core. In this work, we stick to the generous upper limit of 90 oceans in the core from the review by \cite{Peslier2017-SSRvEarthWater}

\paragraph*{Mars:} While Mars is volatile-depleted compared to Earth, its initial water inventory has been largely lost to space through magmatic degassing and atmospheric loss, leaving estimates from 0.014~wt.\% \citep{mccubbinMars2016} to 0.02-0.04~wt.\% \citep{kurokawa2014,jakosky2024} of water in the mantle and the crust. 
However, models of Mars accretion predict early water budget that vary between 0.1-0.2 wt.\% \citep{lunineMars2003, brasser2013,vacher2024}, 0.5 wt.\% \citep{elkins-tantonMars2008,rubie2015}, and up to 2.5-3.4 wt.\% \citep{clesiMars2016}.

\paragraph*{Venus:} The water-equivalent content of Venus is even less constrained than that of Mars because of a substantial lack of data. 
\cite{mccubbin2019} derived a broad range, between $1.2\times 10^{-5}$ and 0.3 wt.\%, for primordial water on Venus. 

\paragraph*{Mercury:} The situation is not better for Mercury. 
\cite{peplowskiMercury2011} suggested that it could have been as water rich as the other terrestrial planets, including Earth. 
However, the oxygen fugacity of Mercury is very low so that, even if it had accreted a substantial amount of water, the latter would likely have been reduced to H$_2$ with an estimated molecular H$_2$O/H$_2$ ratio of 0.034-0.0021 \citep{mccubbin2019}. 
More experimental constraints are needed to better constrain the volatile inventory of Mercury, for example when samples will be available for analysis.


\subsection{Inheritance from interstellar ices}

Figure \ref{fig:terrestrial_water} shows the predicted water-equivalent content as a function of the distance from the Sun in the PSN disk midplane along with the estimates obtained in Mars, Earth, Venus and Mercury, respectively.
The figure reports four curves obtained assuming $T_{1au}=$160, 190, 215 and 225 K, chosen so to cover the upper and lower estimates of Earth's water-equivalent content and the upper limit on Venus. 
In the upper half of the figure, the corresponding four temperature profiles are reproduced.

As in \cite{Boitard-Crepeau2025ApJL}, the snowline, which defines the transition from solid to gaseous water, is not a step function as the condensation assumption \citep[i.e. at 180~K][]{loddersSolarSystemAbundances2003, izidoro2022} would predict, but it is diffuse, with the iced water gradually sublimating across the PSN disk midplane. 
As a result, Earth could have acquired its water content from the rocks in its orbit, not necessitating additional contribution from sources of the outer Solar System.
Actually, the comparison of the water-content curves reported in \cite{Boitard-Crepeau2025ApJL} (their Fig. 2) with the new ones shows the impact of considering the higher BE from the ice first ML in the $T_{1au}$ to match, for example, the minimum water-equivalent content on Earth: $T_{1au}$ is 225 instead of 200 K.
In other words, interstellar grains (i.e., following the MRN grain size distribution of Eq. \ref{eq:MNR}) coated by ice formed during the cold prestellar phase of the PSN could retain enough water even if they coagulated in pebbles when the dust temperature was as high as 225 K.
Once grains coagulated, the trapped ice would be conserved and inherited by the forming planetesimals, asteroids and planets.

On the other hand, the fact that the amount of inherited water (slightly) decreases if one considers a dust distribution with grain sizes 10 times larger (see Fig. \ref{fig:snowlines}) may suggest that the coagulation occurred relatively early, when indeed the temperature in the PSN disk midplane at 1 au was lower than 225 K, corresponding to a solar luminosity of about 0.3--0.5 L$_\odot$ if the dust was thermally heated.

Interestingly, when considering the BE of the ice first layers, our model predicts that also Venus and Mercury may have inherited water from rocks at their respective orbits, even though at lesser extent than Earth which depends on the actual Earth's water content.
For instance, Venus may have acquired up to 2 wt.\% and Mercury 0.1 wt.\% if the Earth's water is the upper estimate by \cite{Peslier2017-SSRvEarthWater}. 
In this case, Mars could also have accreted $\sim9$~wt.\% of water, which is higher than the existing (not well constrained) upper limits (see Sec. \ref{sec:water_planets}).
If Earth's water is in the lowest end of the estimates, then Venus would still have acquired 0.001 wt.\% of water-equivalent content, and Mars would have inherited a bit more than 1~wt.\% of water, consistent with literature estimates. 
When considering the limit by \cite{mccubbin2019} on Venus water-equivalent content, 0.3 wt.\%, the curve passing through it predicts less than 0.001 wt.\% water-equivalent in Mercury.
The same curve would be consistent with the upper limit to the water-equivalent content in Mars and the estimates of Earth.
In this case, the temperature at which grains coagulated in pebbles conserving the water on Earth would be 190 K at the Earth's orbit, 220 K at that of Venus and 290 K at the Mercury's one. 

As discussed in the previous section, the water-equivalent content predictions does not take into account the hydroxylated silicates, obtained by spontaneous deprotonation of water molecules on MgO reactive sites.
Unfortunately, how many water molecules are actually deprotonated by the grains is not easy to evaluate and it is postponed to a future work. 
However, a first rough approximation gives $\sim0.03$~wt.\% of water bound to the grains surface (see discussion in Sec. \ref{subsec:astro-results}). 
While this does not impact the amount of water inherited by Earth, it can largely impact the inherited water content by both Venus and Mercury, as shown in Fig. \ref{fig:terrestrial_water}, as hydroxylated grains can survive even at the Mercury's orbit. 

Finally, as explained in \ref{subsec:inheritance_model}, we assumed $A_{H_2O} = 1\times10^{-4}$ so that our snowline matches the 25~wt.\% water content of the comet 67P at the low temperatures of the outer Solar System. 
Figure \ref{fig:terrestrial_water_appendix} shows the impact of varying the initial water-equivalent amount $A_{H_2O}$ by a factor two (lower and larger) on the curves of Fig. \ref{fig:terrestrial_water}.
As expected, the initial assumed $A_{H_2O}$ impacts the preserved water ice amount at relatively low temperatures, which are dominated by the sublimation of the bulk ice. 
In practice, a factor four uncertainty in the initial water-equivalent amount slightly impacts the predicted predictions for Mars, marginally for Earth (only for the higher value of measured estimates), and does not impact the predictions for Venus and Mercury.

\section{Conclusions}\label{sec:conclusions}

In this work, we modelled the amount of frozen water which envelopes the sub-micron sizes dust grains across the PSN using a kinetic approach and taking into account the BE of the frozen water molecules on the amorphous ice and silicate surfaces, respectively.
For the former, we used the BE distribution computed by \cite{Tinacci2023waterBE}, while for the latter we carried out new QM calculations. 

We found that the water molecules attached to the silicate surface have on average a twice larger BE (about 70~kJ/mol) with respect to those attached to amorphous ice surface.
In addition, we found that water molecules can be chemisorbed on reactive MgO sites on the silicate surfaces, where water is deprotonated, resulting in a very high BE (about 263 kJ/mol) which corresponds to the formation of a chemical covalent bond with the surface.

As in the work by \cite{Boitard-Crepeau2025ApJL}, we show that iced water gradually sublimates from the grain surfaces going inward the PSN, creating a diffuse snowline rather than a sharp desorption front at 180~K, as often assumed by models of the Solar System formation (see Introduction). 
However, with respect to \cite{Boitard-Crepeau2025ApJL}, the addition of the first layers of water molecules attached to the silicate surface increases the temperature at which water can remain frozen on the dust grains.
When comparing our model predictions with the estimates of the water-equivalent content of Earth \citep[e.g.,][]{Peslier2017-SSRvEarthWater}, we found that the sub-micron dust grains at the Earth's orbit, which were covered by ice during the PSN prestellar phase, may have coagulated into the pebbles that eventually formed it when the dust temperature was about 225 K or lower.
This is in agreement with recent observations of coagulation of dust during the Class 0 phase in the solar-type star formation \citep[see discussion in][]{Boitard-Crepeau2025ApJL}.

Remarkably, considering the contribution of the first layers of water ice, attached to the silicate surfaces, leads to predict a water-equivalent content in Mars, Venus and Mercury in agreement with the (poor) estimates that exist.
In the case of Venus and even more of Mercury, the water molecules chemisorbed to the active sites of the silicate surfaces could have contributed to the inheritance of about 0.03 wt.\% of water-content (in their interiors).

Finally, we emphasize once again that this water-equivalent amount would be inherited from local dust grains, initially covered by prestellar ice, without requiring a delivery from Solar System outer bodies. 
This is consistent with sample-based evidences that support the presence of prestellar ice in the inner Solar System \citep[e.g.,][]{delouleRobert1995, piani2015}. 
Likewise, more recently \cite{SossiBower2026}, which are based on the analysis of heavy elements isotopic anomalies, claimed that Earth formed from inner Solar System material.

\section*{Acknowledgements}

Authors acknowledge support from the Project CH4.0 under the MUR program ‘Dipartimenti di Eccellenza 2023-2027’ (CUP: D13C22003520001). 
We acknowledge the EuroHPC Joint Undertaking for awarding this project access to the EuroHPC supercomputer LUMI, hosted by CSC (Finland) and the LUMI consortium through a EuroHPC Regular Access call. 
We acknowledge the CINECA award under the ISCRA initiative, for the availability of high performance computing resources and support.
L. Boitard-Crépeau thanks the Labex OSUG and Université Grenoble Alpes for funding. 
Finally, the authors thank an anonymous referee for insightful comments that helped improving the clarity of the manuscript.

\section*{Data Availability}
The data underlying this article will be shared on reasonable request to the corresponding author.



\bibliographystyle{mnras}
\bibliography{CeciliaCeccarelli} 



\newpage

\clearpage

\appendix

\section{Computational details}\label{apx:comp}

\subsection{Water adsorption strategy}\label{apx:sub_workflow}

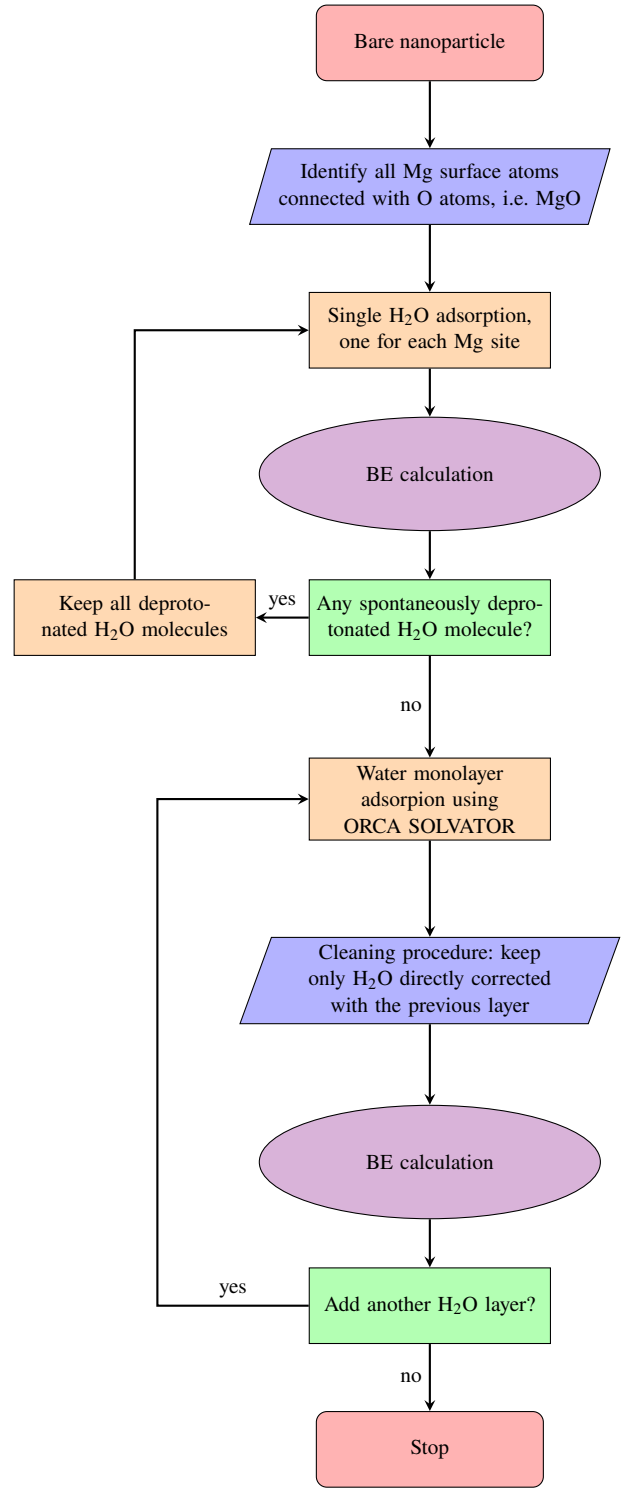
\begin{figure}
    \centering
    \begin{tikzpicture}[node distance=1.4cm]
        \node (start) [startstop] {Bare nanoparticle};
        \node (in1) [io, below of=start, yshift=-0.5cm] {Identify all Mg surface atoms connected with O atoms, i.e. MgO};
        \node (pro1) [process, below of=in1, yshift=-0.5cm] {Single H$_2$O adsorption, one for each Mg site};
        \node (calc1) [calculation, below of=pro1, yshift=-0.5cm] {BE calculation};
        \node (dec1) [decision, below of=calc1, yshift=-0.5cm] {Any spontaneously deprotonated H$_2$O molecule?};        
        \node (pro2b) [process, left of=dec1, xshift=-2.5cm] {Keep all deprotonated H$_2$O molecules};        
        \node (pro2a) [process, below of=dec1, yshift=-1.0cm] {Water monolayer adsorpion using ORCA SOLVATOR};       
        \node (out1) [io, below of=pro2a, yshift=-1.0cm] {Cleaning procedure: keep only H$_2$O directly corrected with the previous layer};       
        \node (calc2) [calculation, below of=out1, yshift=-1.0cm] {BE calculation};        
        \node (dec2) [decision, below of=calc2, yshift=-0.5cm] {Add another H$_2$O layer?};
        \node (stop) [startstop, below of=dec2, yshift=-0.5cm] {Stop};
        
        \draw [arrow] (start) -- (in1);
        \draw [arrow] (in1) -- (pro1);
        \draw [arrow] (pro1) -- (calc1);
        \draw [arrow] (calc1) -- (dec1);
        \draw [arrow] (dec1) -- node[anchor=east] {no} (pro2a);
        \draw [arrow] (dec1) -- node[anchor=south] {yes} (pro2b);
        \draw [arrow] (pro2b) |- (pro1);
        \draw [arrow] (pro2a) -- (out1);
        \draw [arrow] (out1) -- (calc2);
        \draw [arrow] (calc2) -- (dec2);
        \draw [arrow] (dec2.west) -- node[anchor=south] {yes} ++(-2cm,0) -- ++(0,6.7cm) -- (pro2a.west);
        \draw [arrow] (dec2) -- node[anchor=east] {no} (stop);        
    \end{tikzpicture}
    \caption{Flowchart of the $H_2O$ adsorption procedure.}
    \label{fig:workflow}
\end{figure}

\begin{figure*}
    \centering
    \begin{subfigure}[b]{0.48\textwidth}
        \centering
        \includegraphics[width=\textwidth]{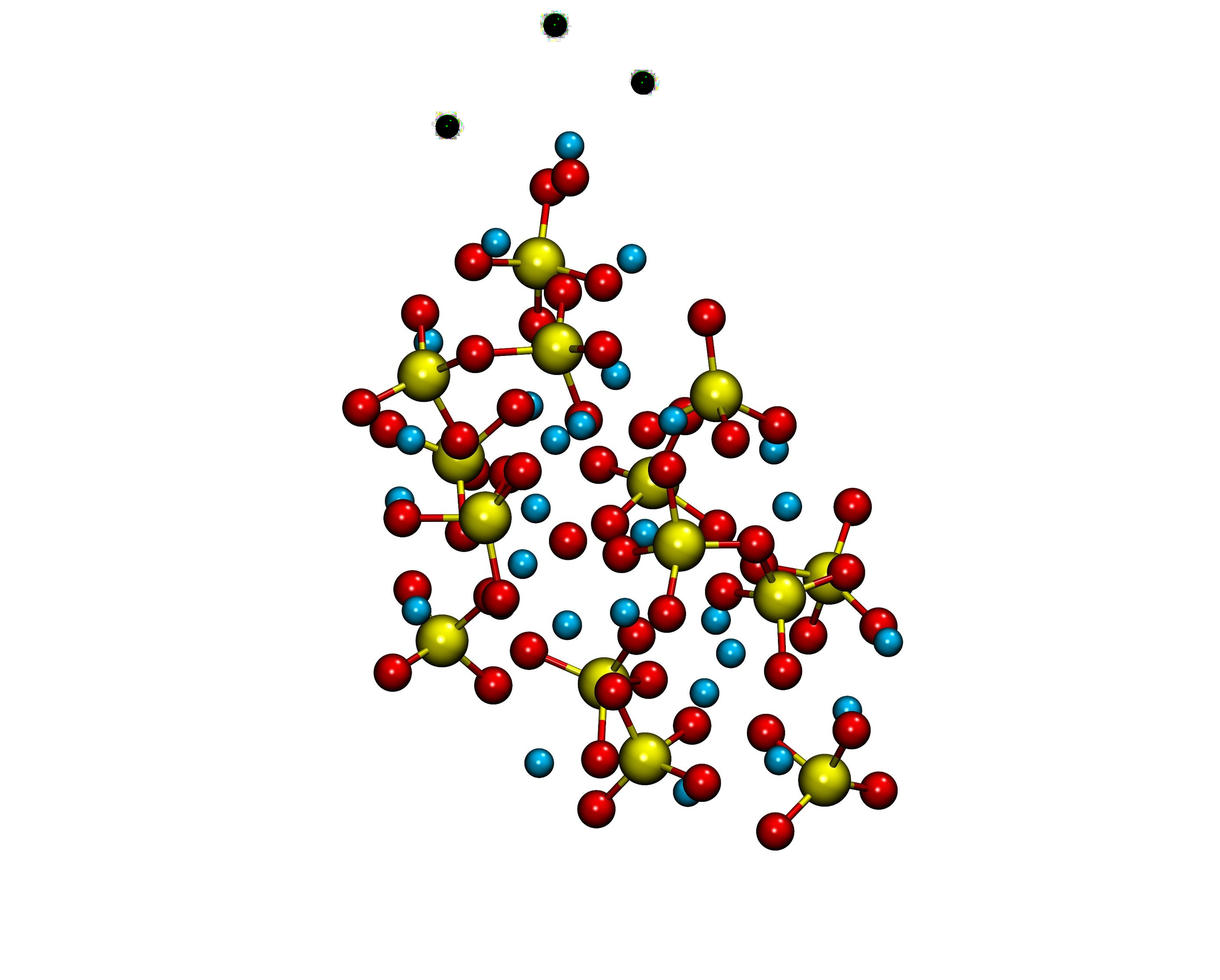}
        \caption{Annealed}
        \label{fig:sub_sites_amrph}
    \end{subfigure}
    \begin{subfigure}[b]{0.48\textwidth}
        \centering
        \includegraphics[width=0.8\textwidth]{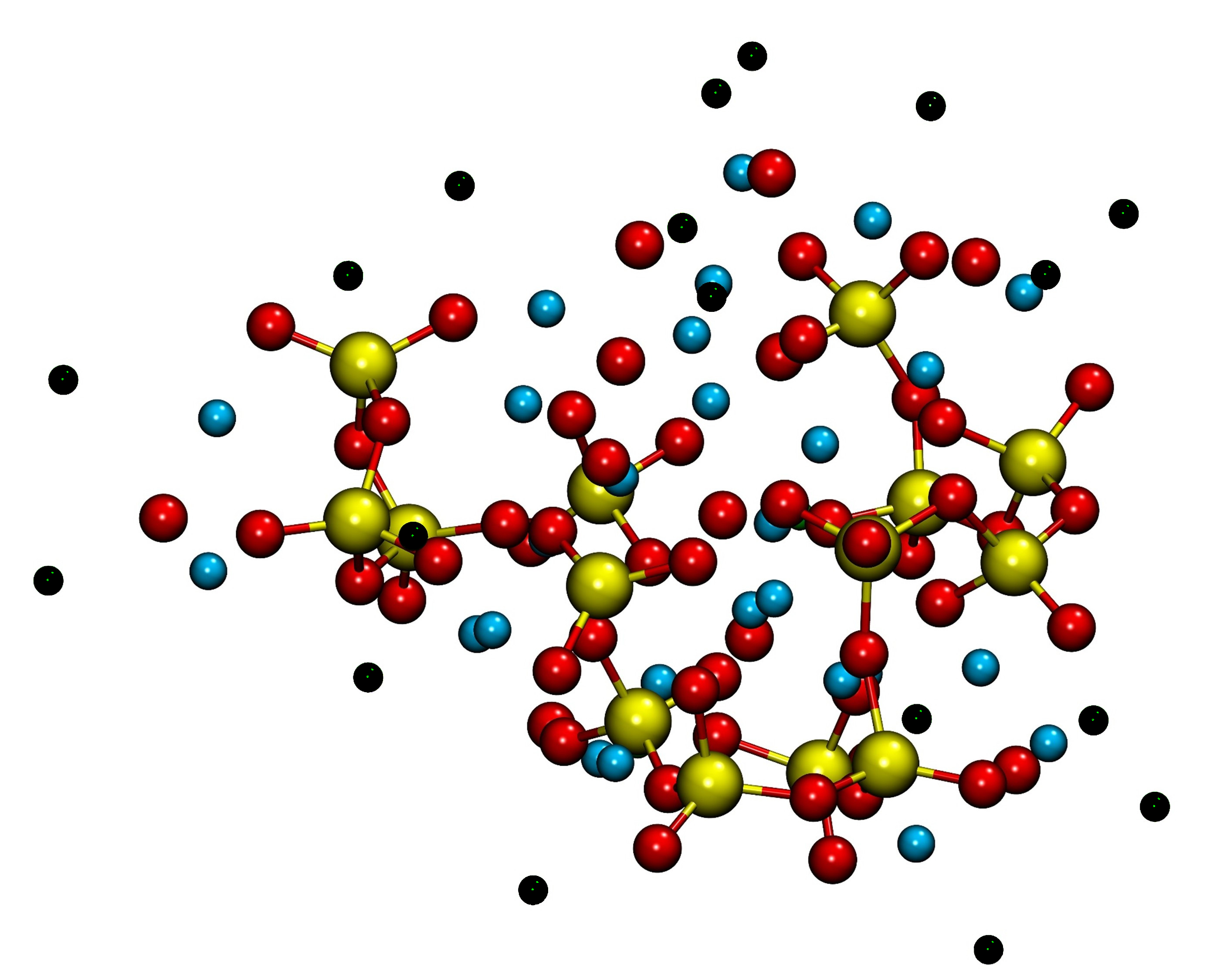}
        \caption{Nucleated}
        \label{fig:sub_sites_ncl}
    \end{subfigure}
    \caption{Water initial binding sites on the outermost Mg atoms. Atom color code: H white, O red, Si yellow, Mg cyan. The black spots represent the center of mass of the water molecule.}
    \label{fig:sites}
\end{figure*}

Fig.~\ref{fig:workflow} shows the flowchart of the H$_2$O adsorption procedure, here expanded in the following steps:
\begin{enumerate}
  \item All surface reactive MgO sites are automatically identified with a Python script, since these sites are the most prone to spontaneous deprotonation of H$_2$O.
  \item At each site a water molecule is placed with the O atom at 2.5~\AA{} from the Mg atom and the H atoms randomly oriented (with the only constrain to point opposite to the center of mass of the nanoparticle) with an O--H distance of 1.0~\AA; the starting positions of the adsorbed H$_2$O are shown in Fig.~\ref{fig:sites}.
  \item Each structure is optimized at r$^2$SCAN-D3(Mg=0) level: a constrained optimization in which only the adsorbed H$_2$O is free to relax, followed by a full optimization of all atoms.
  \item All spontaneously deprotonated H$_2$O are retained for the following simulations, producing a partially hydroxylated nanoparticle.
  \item A second round of single H$_2$O adsorptions was performed only for the nucleated nanoparticle to check whether the hydroxylated nanoparticle can further deprotonate H$_2$O that were stable in their molecular form in the first round.
  \item The hydroxylated nanoparticle is then completely covered by a H$_2$O monolayer using the SOLVATOR tool (ORCA) \citep{Neese2025} in combination with the GFN-FF pre-sampling \citep{Spicher2020}, freezing all atoms of the forsterite nanoparticle, to avoid unpredictable structural collapse due to weakness of the GFN-FF parametrization for inorganic materials.
  \item To ensure a monolayer coverage, 100 water molecules were initially placed; after DFT geometry optimization all H$_2$O not directly linked to forsterite atoms (deprotonated H$_2$O being considered as part of forsterite) were removed. For a detailed description on the cleaning procedure see Appendix \ref{apx:water_conn}.
  \item Steps vi–vii were repeated for the second and third water layers, for which 200 and 500 H$_2$O were initially added, respectively, followed by DFT geometry optimization and removal of excess waters.
\end{enumerate}


\subsection{Water layer definition: connectivity and cleaning procedure}\label{apx:water_conn}


To simulate the multilayer, and in particular to calculate the binding energy associated with each layer, it is important to define the meaning of a single water layer. In the specific case of the hydroxylated nanoparticles, the first water layer (monolayer) is defined by all water molecules in direct contact with forsterite atoms (including those belonging to chemisorbed waters). The other water layers follow the same principle: the second layer is defined by the water molecules in direct contact with those belonging to the monolayer, and so on.

\begin{figure*}
    \centering
    \begin{subfigure}[b]{0.48\textwidth}
        \centering
        \includegraphics[width=\textwidth]{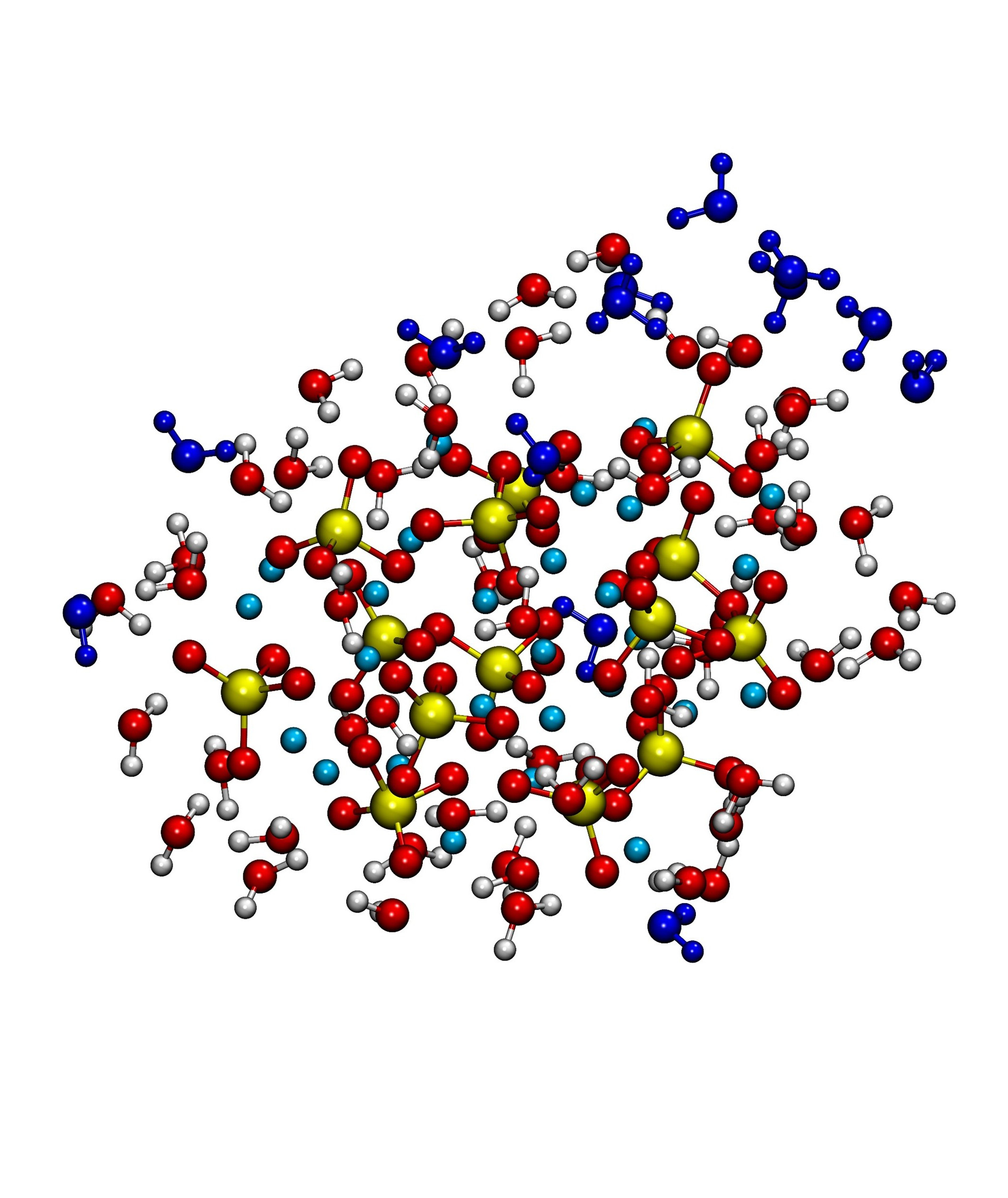}
        \caption{Annealed}
        \label{fig:sub_amrph_h2o_mono}
    \end{subfigure}
    \begin{subfigure}[b]{0.48\textwidth}
        \centering
        \includegraphics[width=\textwidth]{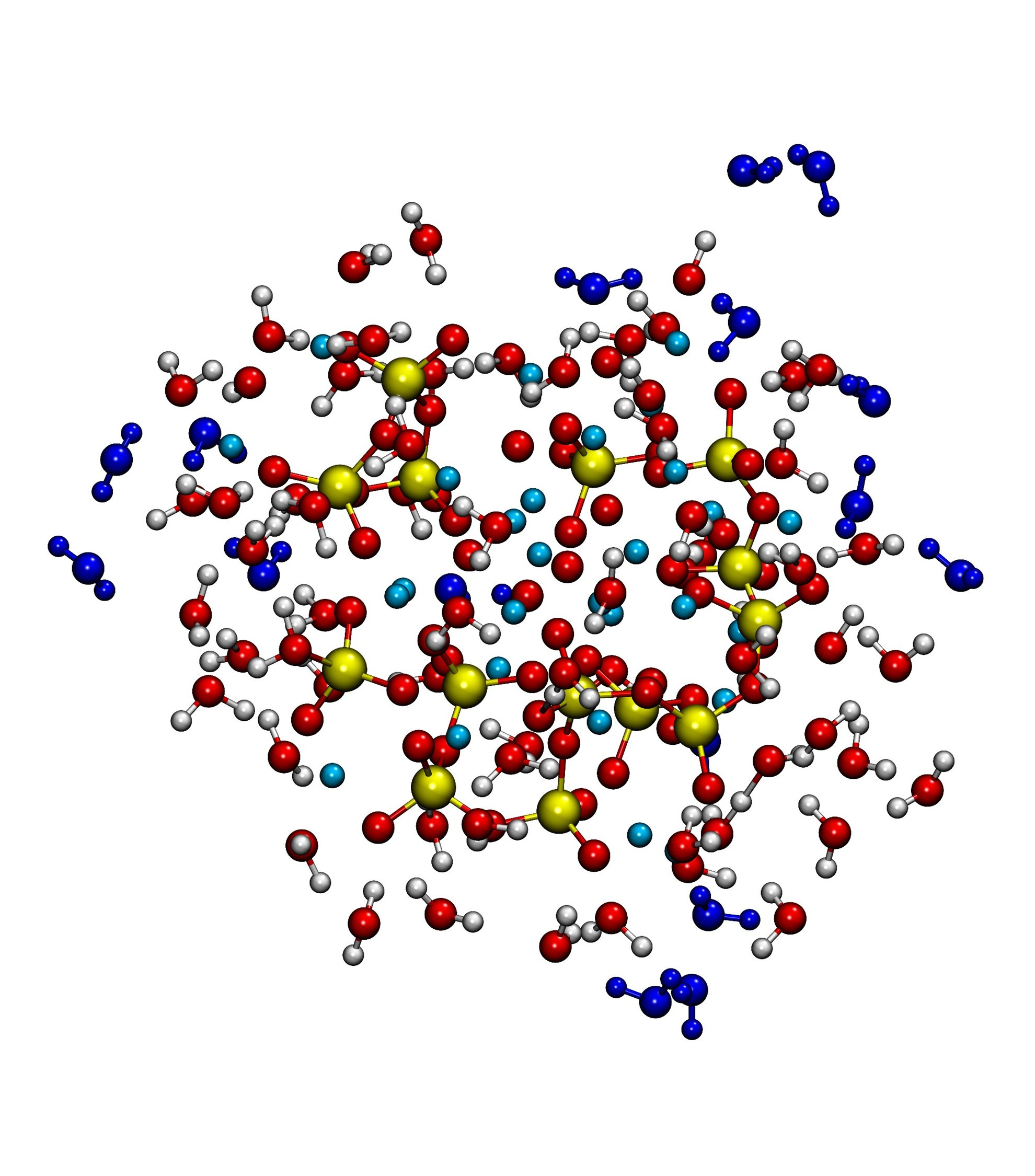}
        \caption{Nucleated}
        \label{fig:sub_ncl_h2o_mono}
    \end{subfigure}
    \caption{$H_2O$ monolayer on the annealed and nucleated nanoparticles. The figure highlights the two criteria used for the ASE covalent radii modifications (+1.0 \AA{} for red and white waters and +1.5 \AA{} for blue waters) of the H atom in order to properly include H-bond interactions. Atom colour code: H white, O red, Si yellow, Mg cyan.}
    \label{fig:h2o_monolayer}
\end{figure*}

When we simulated the adsorption of the water monolayer with the ORCA SOLVATOR, we docked an excess of water molecules (100). Then, a cleaning procedure was applied in order to reach to proper number of water molecules which define the monolayer, according to the previously mentioned principle.
To define the contacts among water molecules we used the connectivity based on interatomic distances as defined by tabulated covalent radii, which had to be modified in order to properly catch intermolecular H-bond interactions.
To this end, the default parameters for the covalent radii definition as implemented in the Atomistic Simulation Environment (ASE) were used, apart for the H atom, whose covalent radius (0.31~\AA) was increased of 1.0 and 1.5~\AA.
Fig.~\ref{fig:h2o_monolayer} shows the difference between the two criteria used for the H-bond definition. As one can see from a visual point of view, the 1.0 criterion better represents the above-mentioned definition of monolayer, indeed the majority of H$_2$O highlighted in blue are connected with other H$_2$O molecules, and not directly with forsterite atoms.
Therefore, to build up successive H$_2$O layers only the 1.0 criterion was used, even if, once the second (and third) layers were adsorbed, both criteria were used to remove the water molecules in excess.







\clearpage

\twocolumn
\section{Results}\label{apx:results}
\begin{table}
    \centering
    \caption{Experimental (Exp) and computed (Comp) bulk structure and sublimation enthalpy of the crystalline hexagonal ice. 
    The experimental values are obtained on Ih ice (proton disordered), while computed  ones on the P-ice (proton ordered). Cell parameters are in \AA, cell angles in degrees, and cell volumes in \AA$^3$. Energy in~kJ~mol$^{-1}$.}
    \label{tab:ice_data}
    \begin{tabular}{lccccc}
        \hline
&  a & b & $\gamma$ & V & $\Delta H^{sub}$ \\
        \hline
Exp	& 4.381$^a$	& 7.183$^a$ &	120.000$^a$ & 	119.421$^a$ &	 -59.2$^b$\\
Comp &	4.378 & 	7.180	& 120.086 &	119.222	& -57.5\\
$\Delta$\%	& 0.05	& 0.03	& -0.07	& 0.17	& -2.96\\
        \hline
    \multicolumn{6}{l}{$^a$\citet{Chodkiewicz2024}}\\
    \multicolumn{6}{l}{$^b$\citet{chickos2002enthalpies}}\\
    \end{tabular}
\end{table}
\subsection{Accuracy of the method}\label{apx:sub_comp_exp}

In Table~\ref{tab:ice_data} we report the experimental and computed cell parameters for the hexagonal (Ih) and proton ordered (P-) water ices, as well as the sublimation enthalpy. The latter is calculated as follows:

\begin{equation}
    \Delta H^{sub} = \frac{H_{ice} - n H_{H_2O}}{n}
\end{equation}

\noindent where $H_{ice}$ and $H_{H_2O}$ are the enthalpies, i.e. the electronic energy + zero-point energy + thermal corrections calculated at 298 K, of P-ice and the isolated $H_2O$ molecule. It is worth noticing that experimentally the ice is proton disordered, while in the computed case it is proton ordered (P-ice). However, it was previously demonstrated the accuracy of using the P-ice model, on several different properties \citep{pisani1996proton}. Indeed, both cell parameters and sublimation enthalpies are very well represented.

\subsection{Single $H_2O$ adsorption}\label{apx:sub_h2o_ads}

The first step of the H$_2$O adsorption procedure was the identification of the MgO defective sites on the surface of both the annealed and nucleated nanoparticles. Previous experimental work confirm the trend of increasing reactivity from crystalline to amorphous forsterite, to MgO, possibly due to the greater basicity of O$^{2-}$ compared to SiO$_4^{4-}$ anions \citep{mates2025revealing}. 

In Fig.~\ref{fig:sites} all the starting H$_2$O adsorption geometries are reported as black spots atop the annealed and nucleated nanoparticles. As one can see, on the annealed nanoparticle there is only one isolated O$^{2-}$ anion, surrounded by three Mg$^{2+}$ cations, which are the only three adsorption sites explored. Only one of them does not deprotonate H$_2$O, with BE = 98.0~kJ~mol$^{-1}$, while on the other two H$_2$O underwent spontaneous deprotonation, with BE = 186.5 and 262.7~kJ~mol$^{-1}$. The most stable one was taken to proceed with the monolayer adsorption.

\begin{figure}
    \centering
    \begin{subfigure}[b]{0.33\textwidth}
        \centering
        \includegraphics[width=\textwidth]{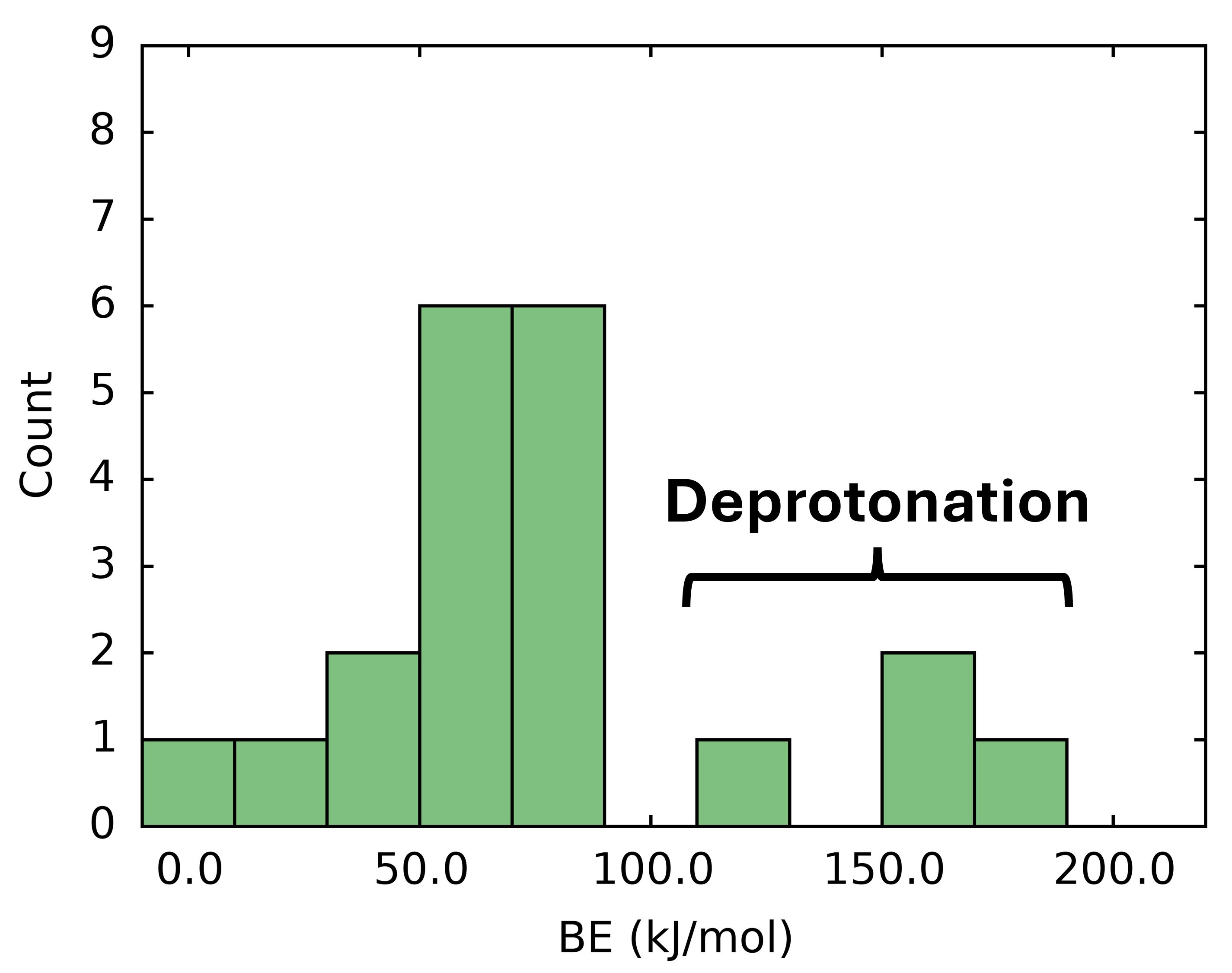}
        \caption{First round}
        \label{fig:sub_distribution1}
    \end{subfigure}
    \begin{subfigure}[b]{0.33\textwidth}
        \centering
        \includegraphics[width=\textwidth]{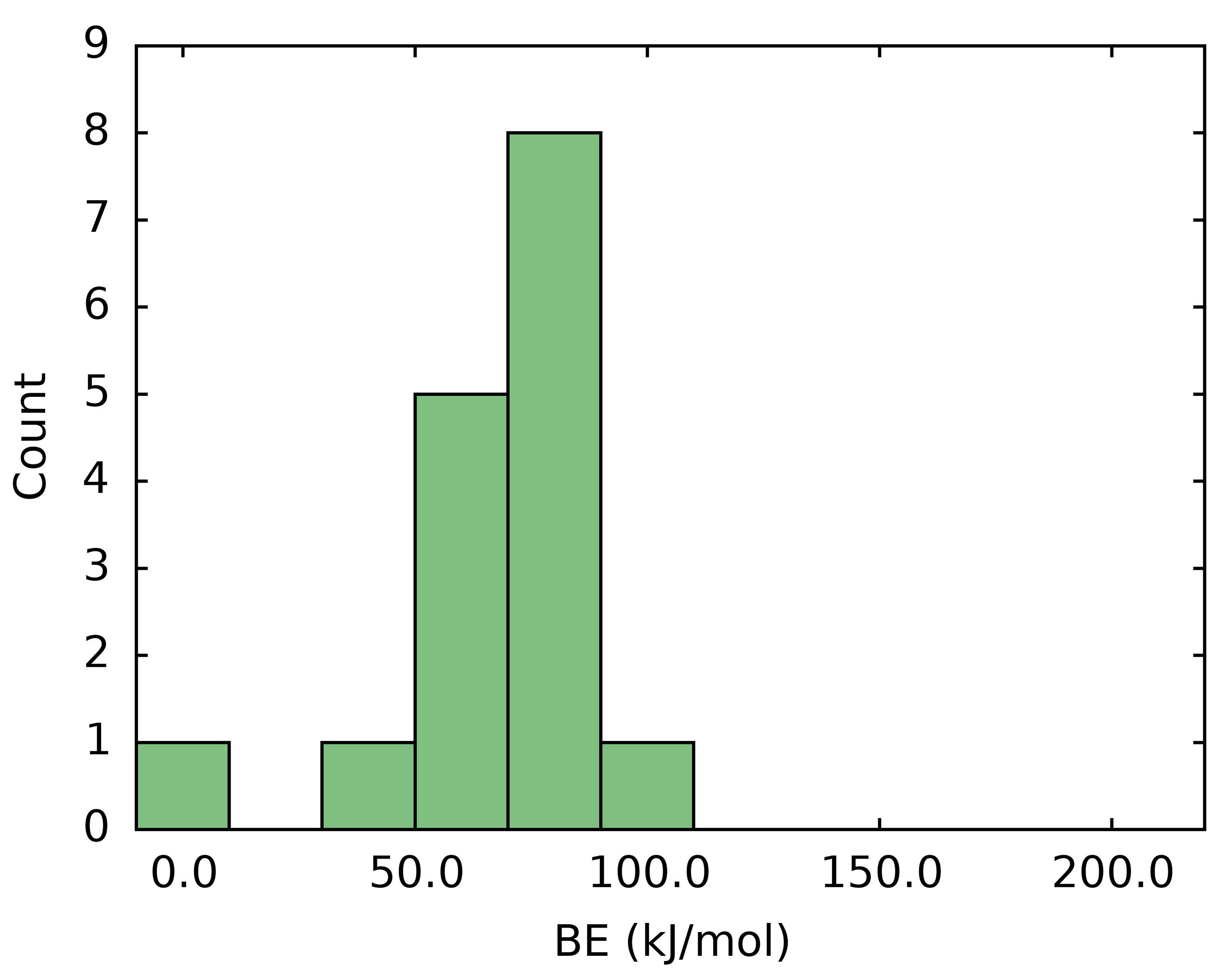}
        \caption{Second round}
        \label{fig:sub_distribution2}
    \end{subfigure}
    \caption{Single H$_2$O binding energy distribution on the nucleated nanoparticle. The first round is obtained adsorbing H$_2$O on each Mg site of the bare nanoparticle. The second round is the same, but starting from the hydroxylated nanoparticle  obtained during the first round.}
    \label{fig:distribution_h2o}
\end{figure}

\begin{figure*}
    \centering
    \begin{subfigure}[b]{0.3\textwidth}
        \centering
        \includegraphics[width=\textwidth]{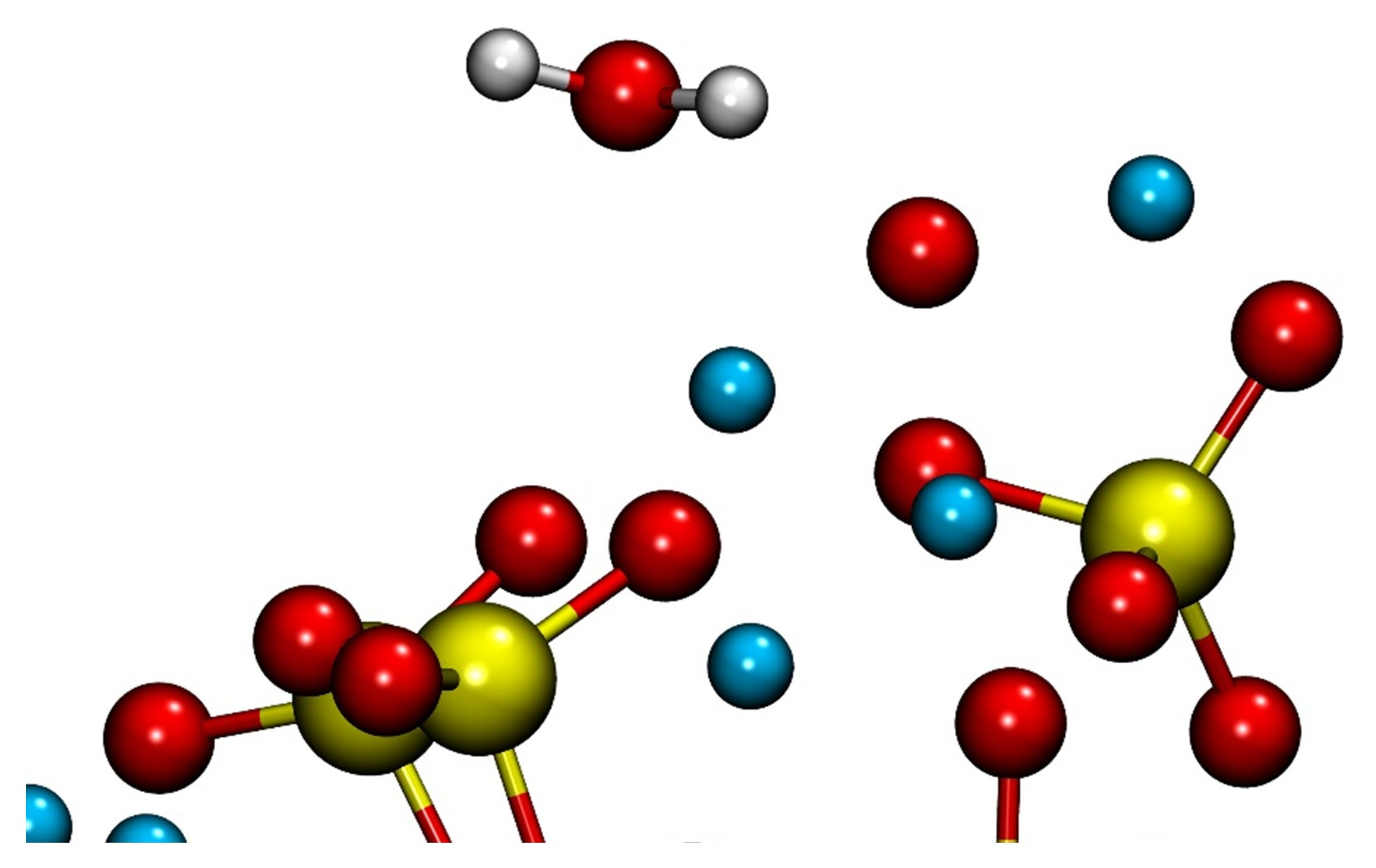}
        \caption{Annealed 1W – start}
        \label{fig:sub_amrph_start}
    \end{subfigure}
    \begin{subfigure}[b]{0.3\textwidth}
        \centering
        \includegraphics[width=\textwidth]{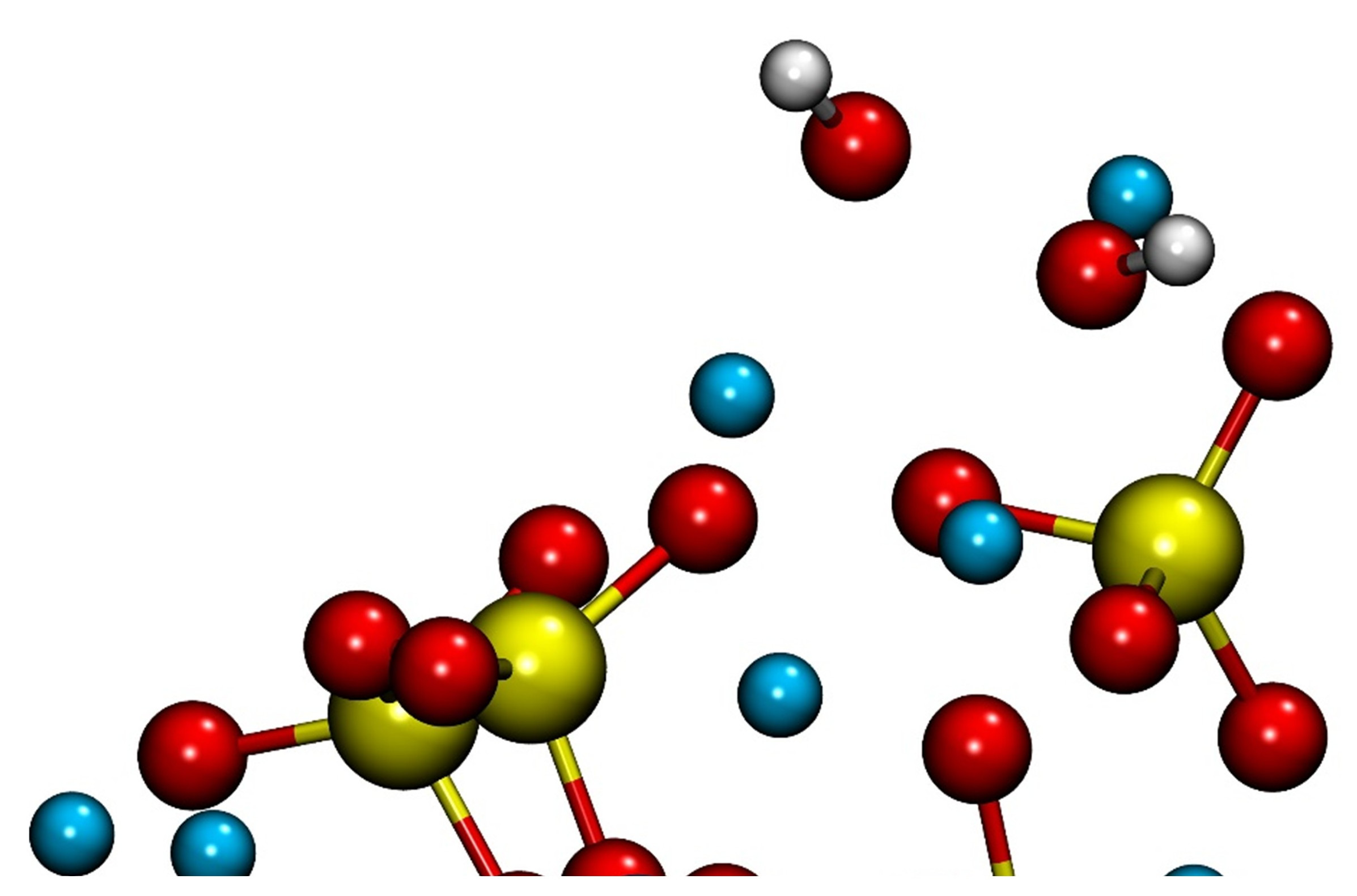}
        \caption{Annealed 1W – opt (BE = 262.7)}
        \label{fig:sub_amrph_opt}
    \end{subfigure}\\
    
    \begin{subfigure}[b]{0.3\textwidth}
        \centering
        \includegraphics[width=0.8\textwidth]{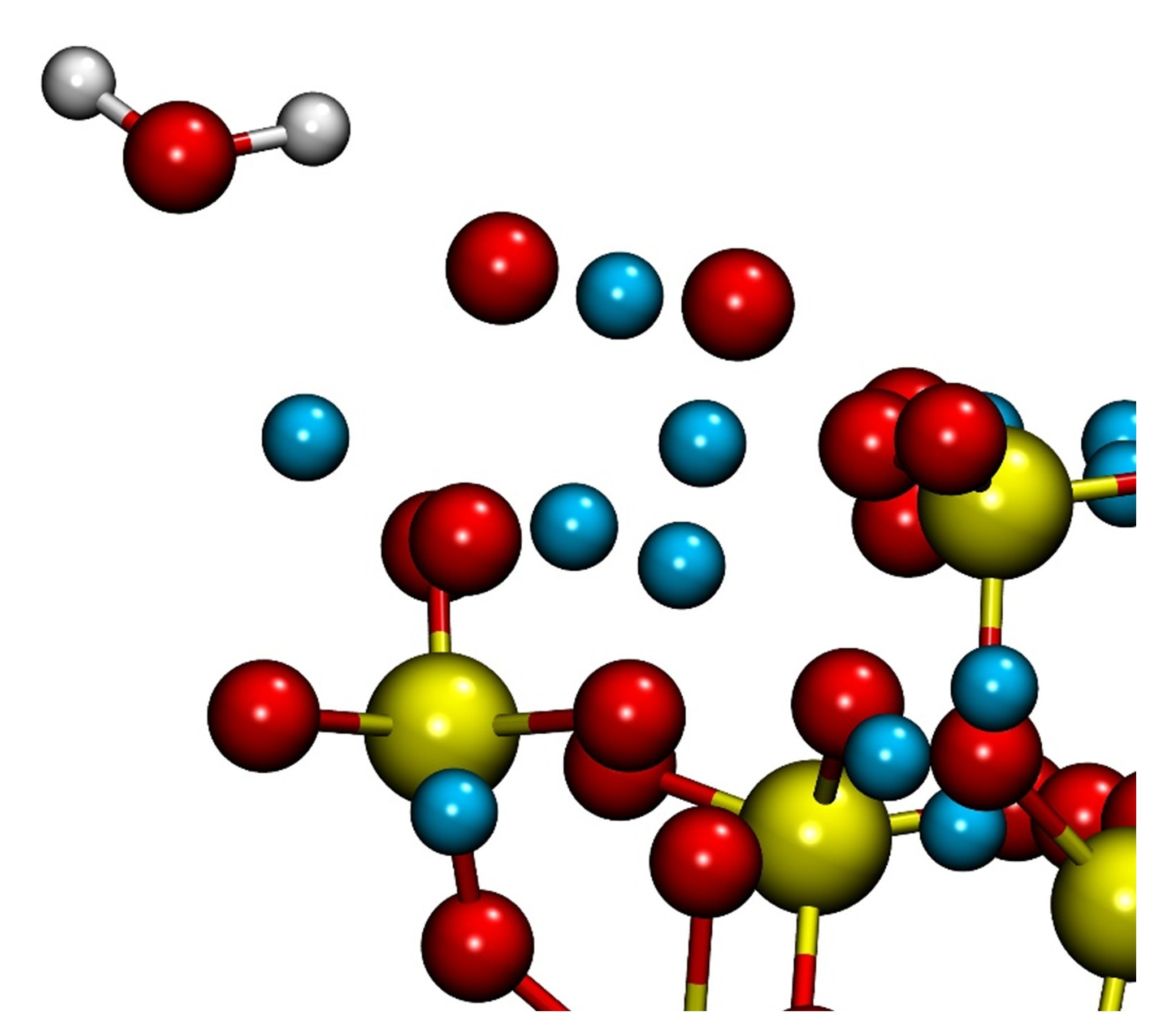}
        \caption{Nucleated 1W – start}
        \label{fig:sub_ncl_1_start}
    \end{subfigure}
    \begin{subfigure}[b]{0.3\textwidth}
        \centering
        \includegraphics[width=0.8\textwidth]{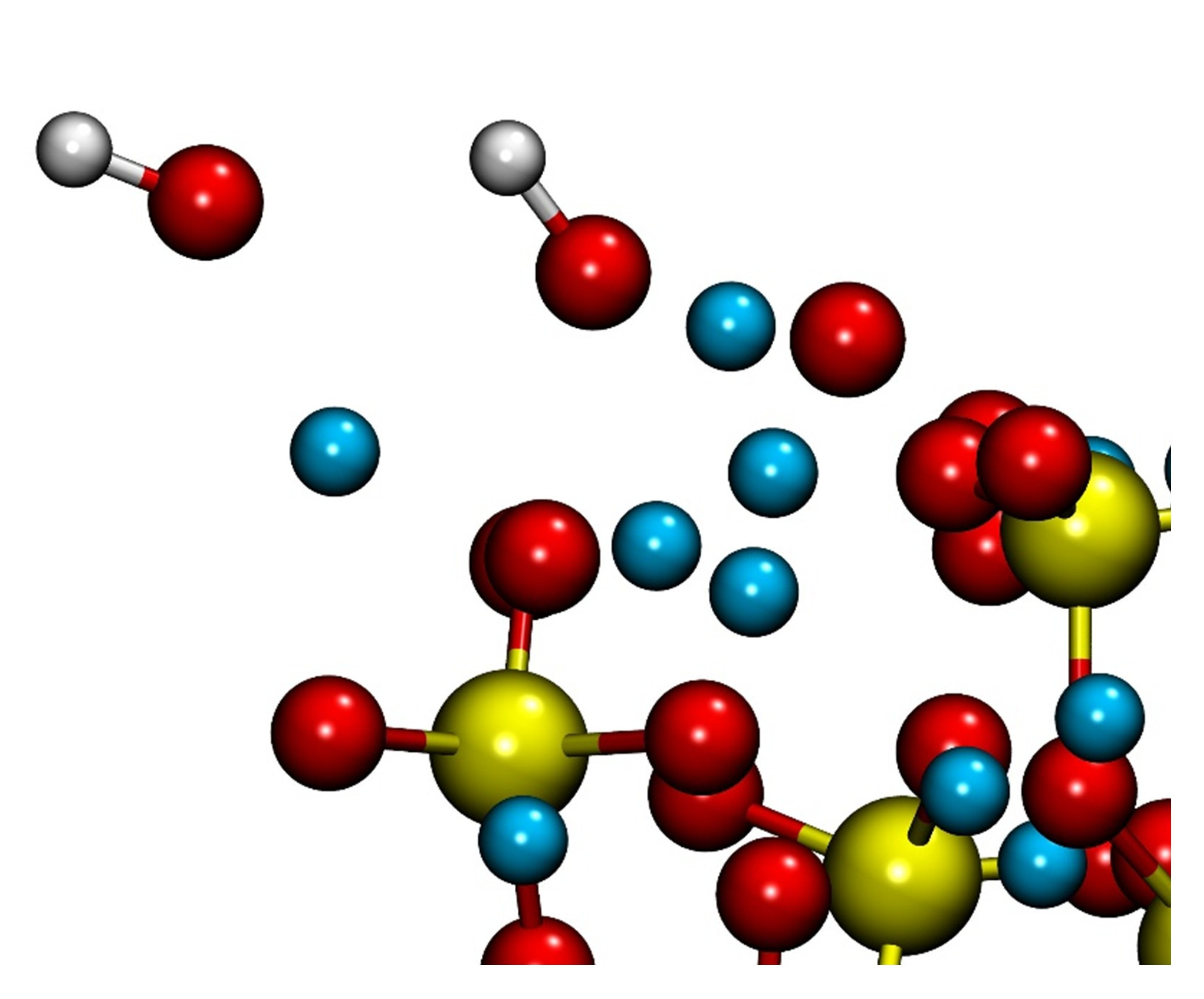}
        \caption{Nucleated 1W – opt (BE = 139.2)}
        \label{fig:sub_ncl_1_opt}
    \end{subfigure}\\
    
    \begin{subfigure}[b]{0.3\textwidth}
        \centering
        \includegraphics[width=\textwidth]{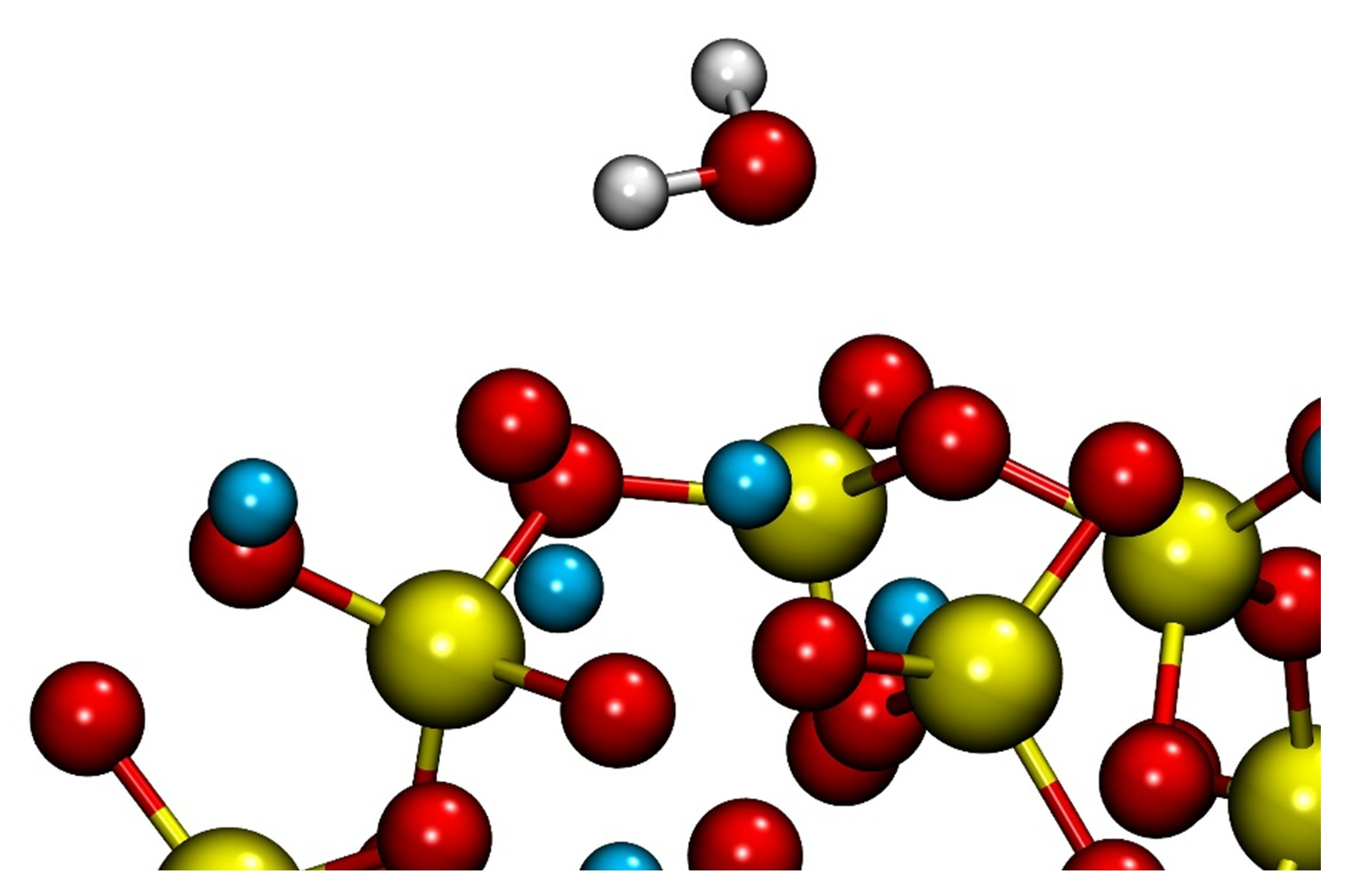}
        \caption{Nucleated 2W – start}
        \label{fig:sub_ncl_2_start}
    \end{subfigure}
    \begin{subfigure}[b]{0.3\textwidth}
        \centering
        \includegraphics[width=\textwidth]{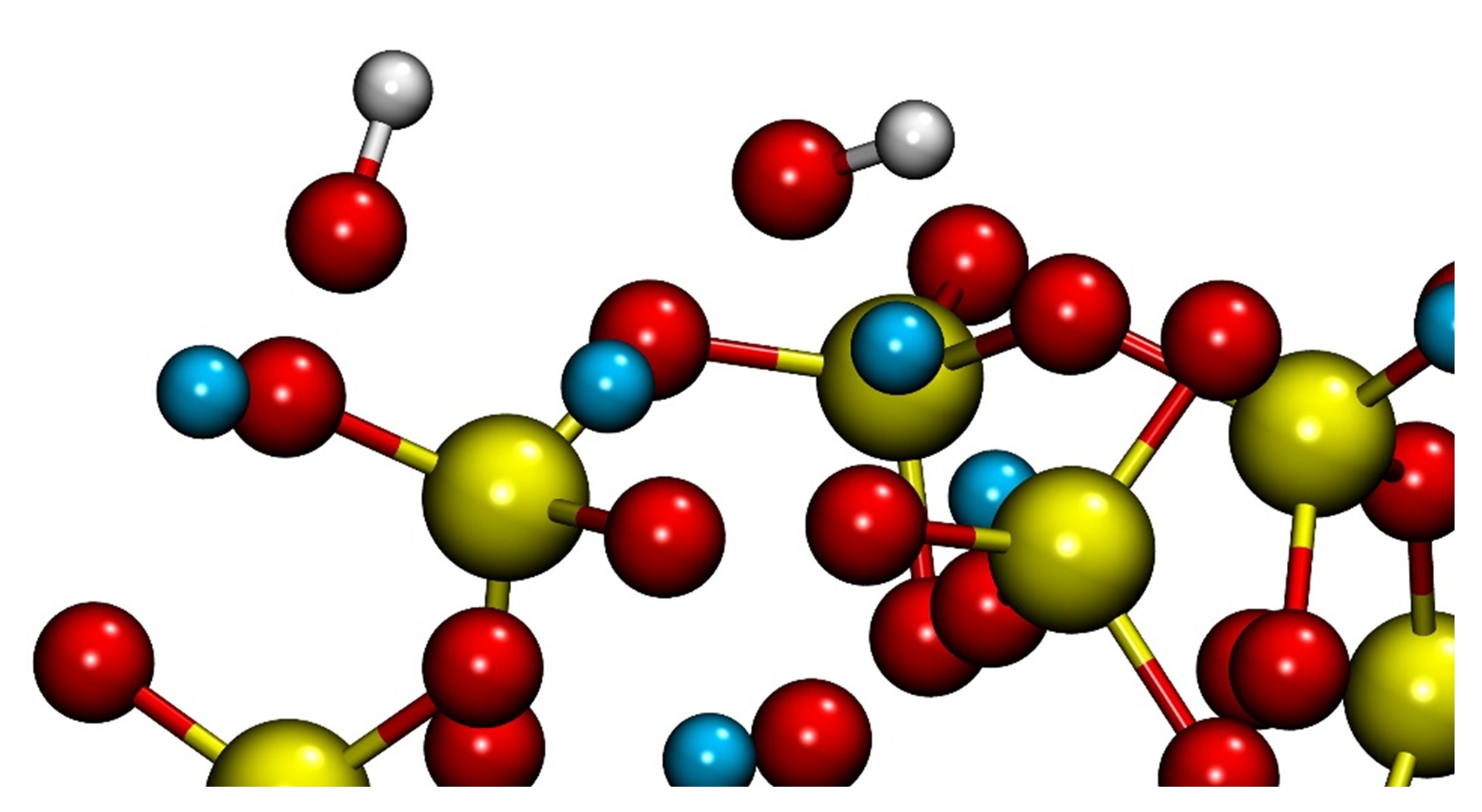}
        \caption{Nucleated 2W – opt (BE = 166.6)}
        \label{fig:sub_ncl_2_opt}
    \end{subfigure}\\
    
    \begin{subfigure}[b]{0.3\textwidth}
        \centering
        \includegraphics[width=\textwidth]{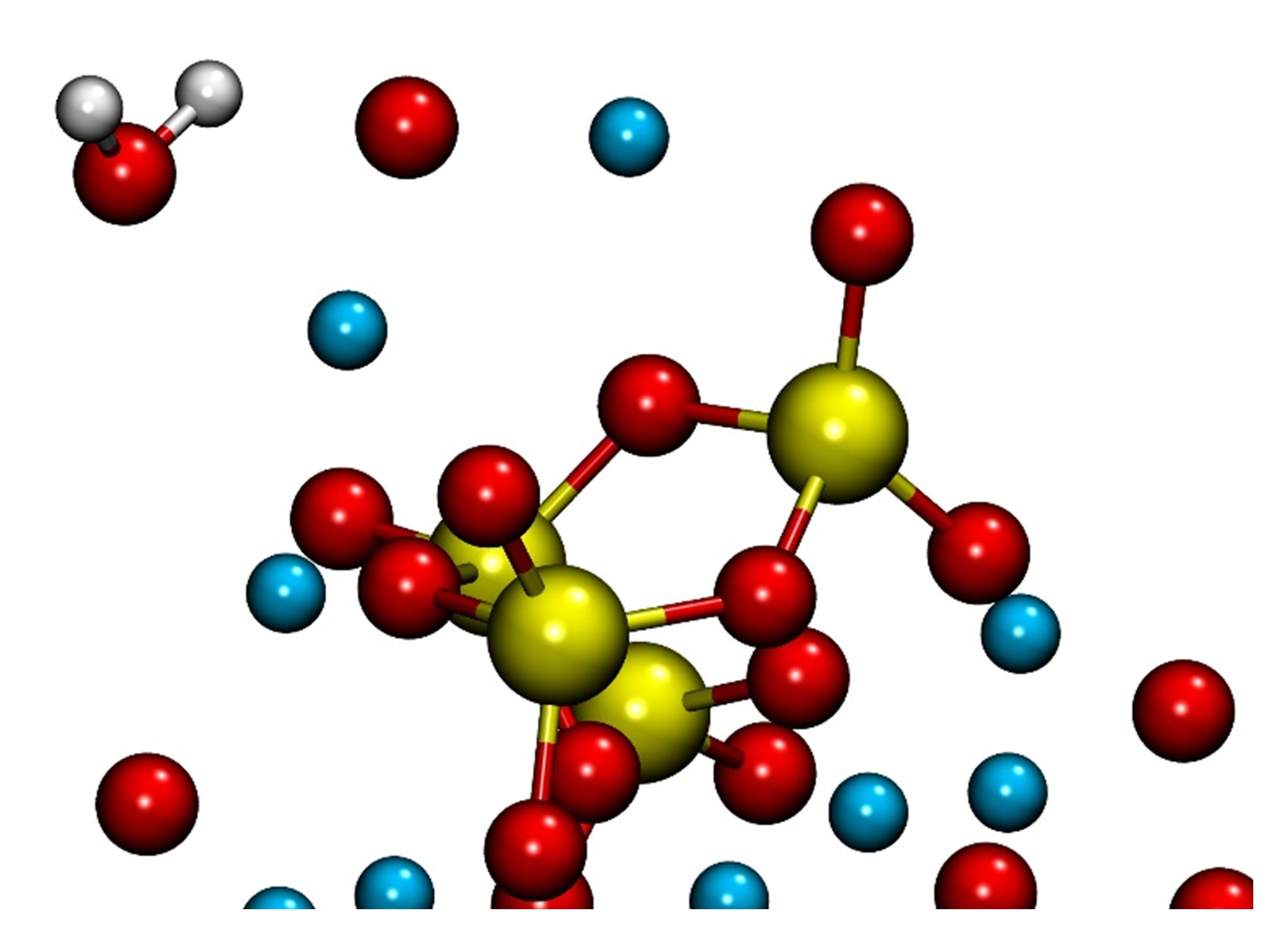}
        \caption{Nucleated 3W – start}
        \label{fig:sub_ncl_3_start}
    \end{subfigure}
    \begin{subfigure}[b]{0.3\textwidth}
        \centering
        \includegraphics[width=\textwidth]{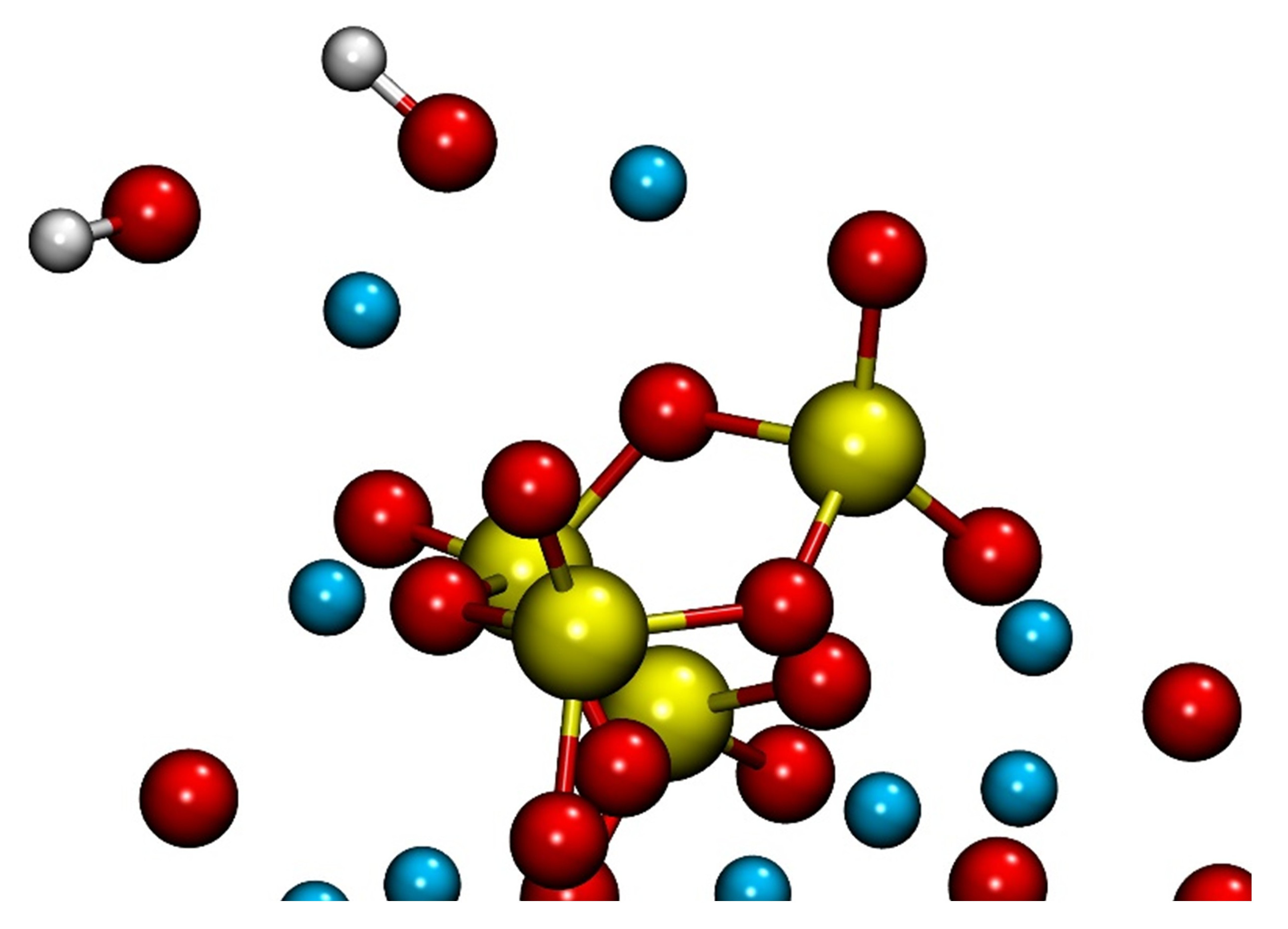}
        \caption{Nucleated 3W – opt (BE = 170.1)}
        \label{fig:sub_ncl_3_opt}
    \end{subfigure} \\
    
    \begin{subfigure}[b]{0.3\textwidth}
        \centering
        \includegraphics[width=\textwidth]{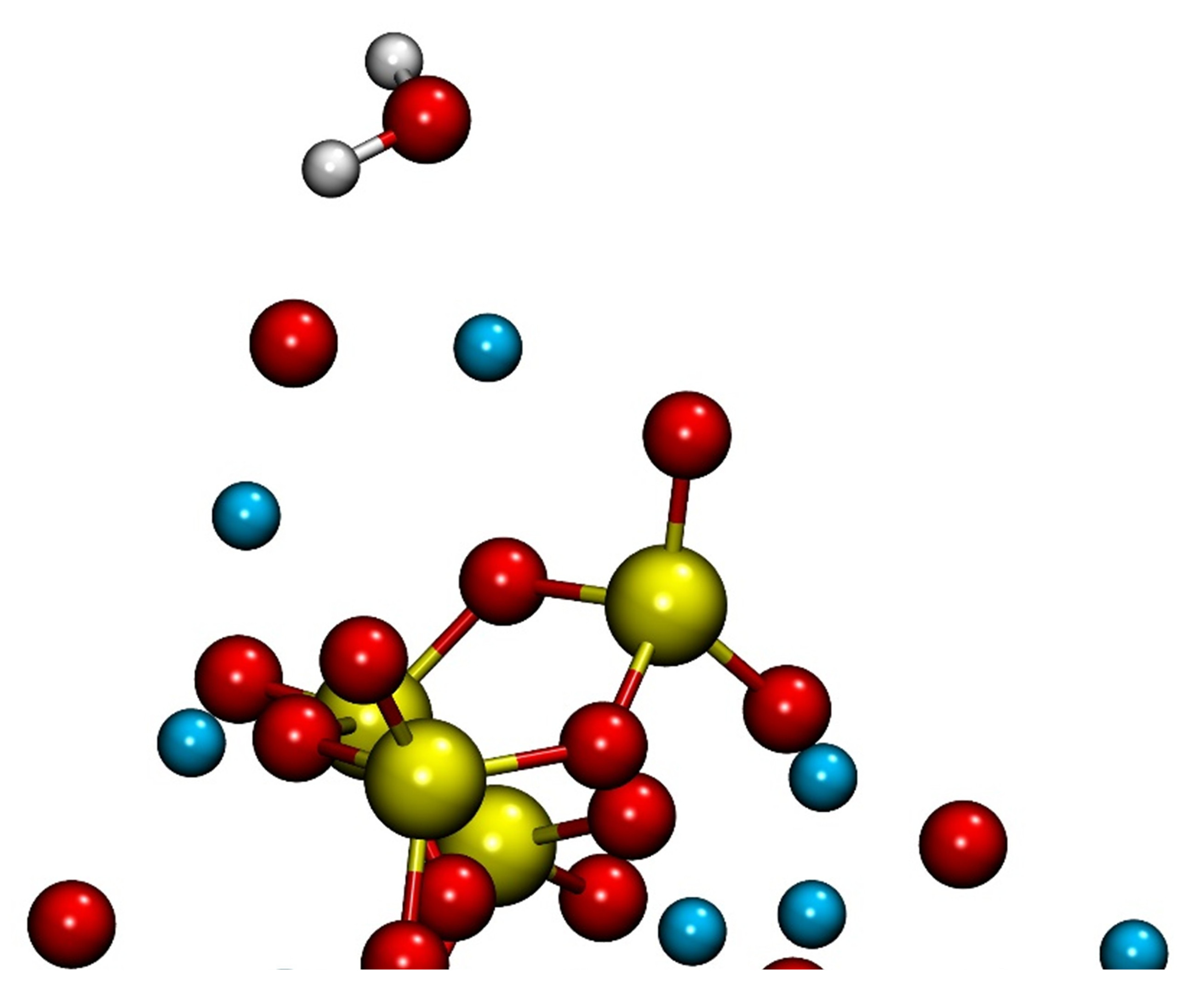}
        \caption{Nucleated 4W – start}
        \label{fig:sub_ncl_4_start}
    \end{subfigure}
    \begin{subfigure}[b]{0.3\textwidth}
        \centering
        \includegraphics[width=\textwidth]{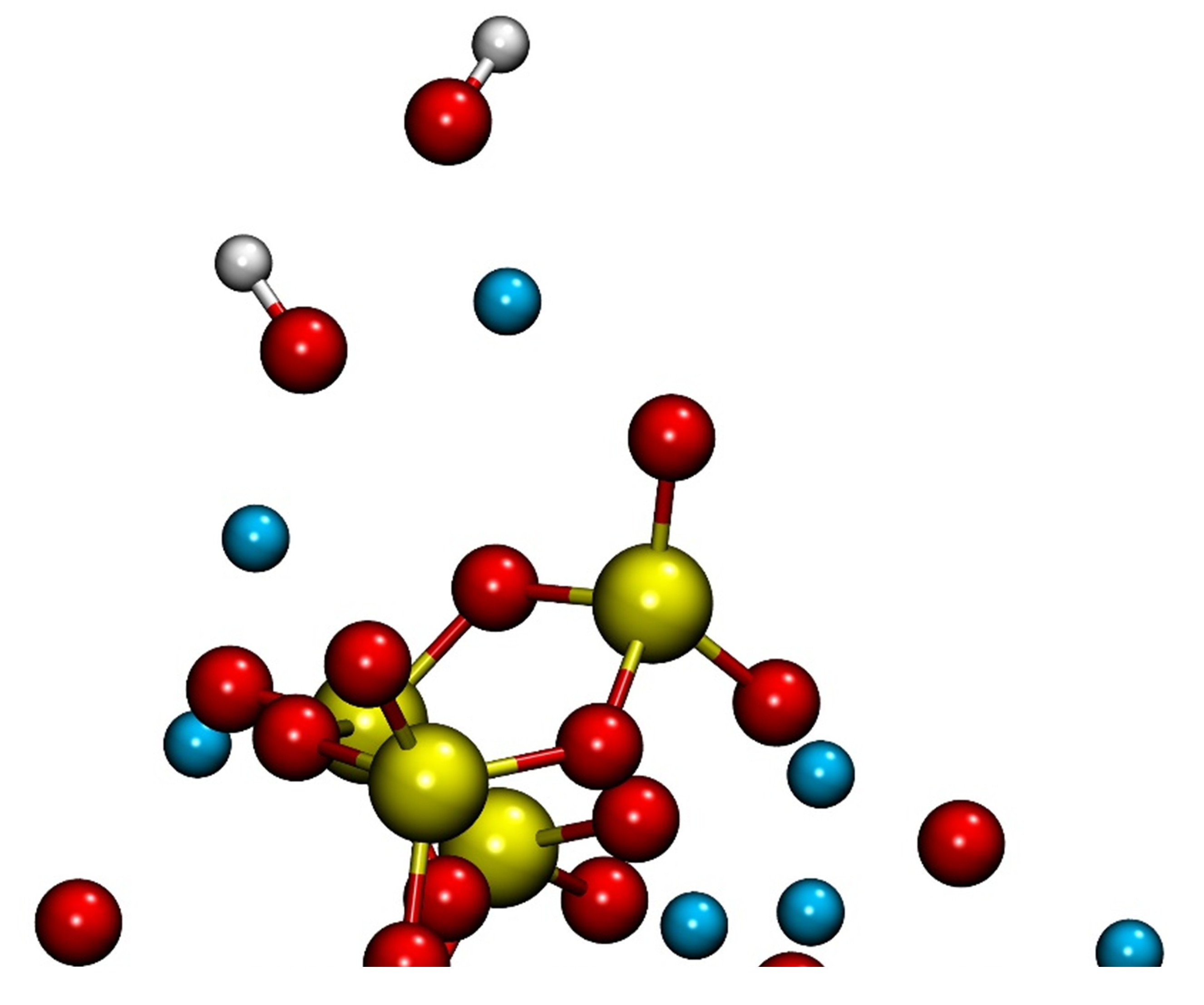}
        \caption{Nucleated 4W – opt (BE = 185.8)}
        \label{fig:sub_ncl_4_opt}
    \end{subfigure}
    \caption{Initial and optimised (r$^2$SCAN-D3(Mg=0)/TZVP) structures of the spontaneous deprotonation sites on the annealed and nucleated nanoparticles. Atom colour code: H white, O red, Si yellow, Mg cyan.}
    \label{fig:struc_start_opt}
\end{figure*}

As regards the nucleated nanoparticle, we carried out a first round exploring all available MgO binding sites (20). In the first round, H$_2$O deprotonates on four different Mg sites (see Figs.~\ref{fig:distribution_h2o} and  \ref{fig:struc_start_opt}); however, as in two cases the H$^+$ goes on the same O (Figs.~\ref{fig:sub_ncl_3_opt} and ~\ref{fig:sub_ncl_4_opt}), only the most stable case was taken for successive adsorptions (i.e. Fig.~\ref{fig:sub_ncl_4_opt}), thus obtaining a total number of three spontaneously deprotonated H$_2$O molecules. The average binding energy of the above-mentioned H$_2$O molecules is BE = 179.4~kJ~mol$^{-1}$ (see Table~\ref{tab:BE_table}).

\begin{figure*}
    \centering
    \begin{subfigure}[b]{0.45\textwidth}
        \centering
        \includegraphics[width=0.7\textwidth]{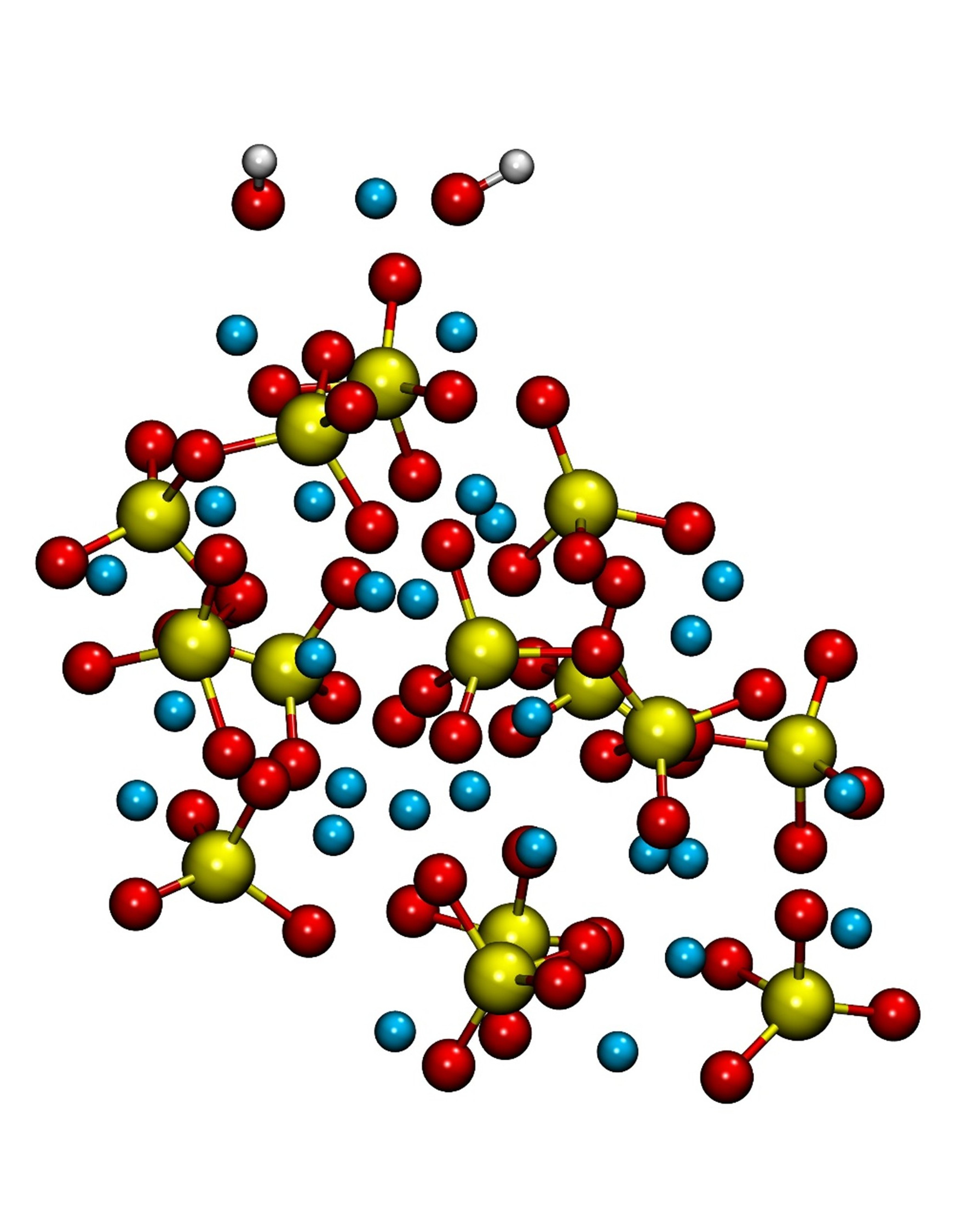}
        \caption{Annealed}
        \label{fig:sub_amrph_h2o_chem}
    \end{subfigure}
    \begin{subfigure}[b]{0.45\textwidth}
        \centering
        \includegraphics[width=0.9\textwidth]{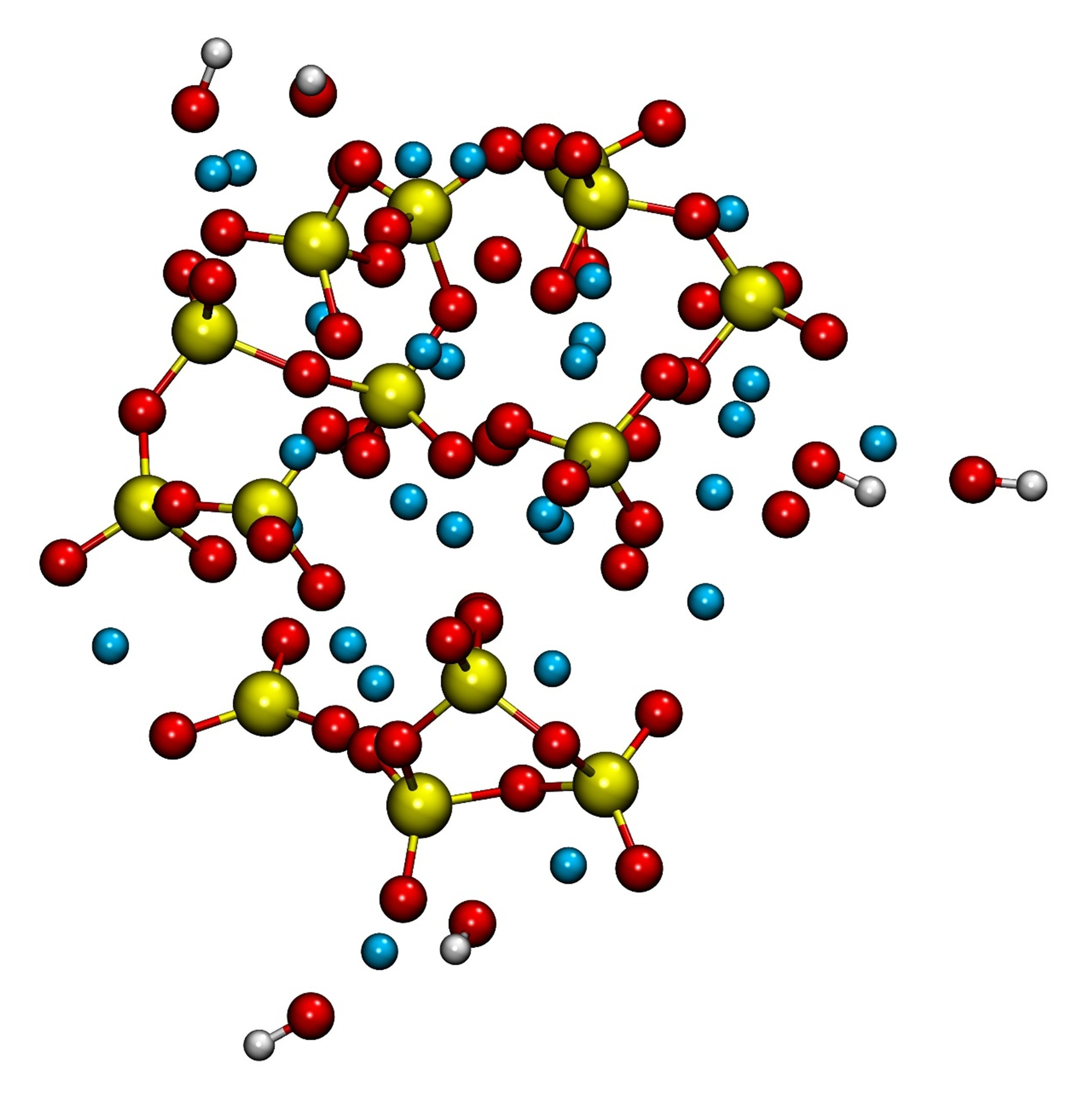}
        \caption{Nucleated}
        \label{fig:sub_ncl_h2o_chem}
    \end{subfigure}
    \caption{Optimised structure of the Annealed and Nucleated hydroxylated nanoparticles after spontaneous deprotonation of water. Atom colour code: H white, O red, Si yellow, Mg cyan.}
    \label{fig:h2o_chemisorbed}
\end{figure*}

A second round of adsorption was carried out using as reference structure the hydroxylated nucleated nanoparticle which does not reveal any further spontaneous deprotonation, confirmed by both structural and energetic features. This also confirms that forsterite, or better, the SiO$_4^{4-}$ units are not so prone to spontaneously deprotonate H$_2$O compared to O$^{2-}$. Indeed, spontaneous deprotonations (chemisorptions) always lead to much stronger binding energies, almost twice as large as that of physisorptions. The final structures of the annealed and nucleated hydroxylated nanoparticles are reported in Fig.~\ref{fig:h2o_chemisorbed}.

\subsection{$H_2O$ multilayer}\label{apx:sub_h2o_multi}

Table~\ref{tab:appendix_BE_ML} reports all the most important information about the H$_2$O coverage of the forsterite nanoparticles: the number of water molecules included in the simulation, the binding energy (BE) and the dipole moment of the nanoparticles, before and after simulating the H$_2$O mantle. The total number of H$_2$O molecules ($nW_{tot}$) includes chemisorbed waters ($nW_{OH}$), as discussed in the previous section, i.e. 1 and 3 for the annealed and nucleated nanoparticles, and the waters added with ORCA (after automatic cleaning procedure), for each layer considered, up to the third ML  ($nW_{ML1}$, $nW_{ML2}$, and $nW_{ML3}$). All BE of physisorbed water fairly below the BE of chemisorbed (i.e. deprotonated) waters (Fig.~\ref{fig:sub_distribution1}), and also below the upper limit of the binding energy distribution of physisorbed waters, i.e. the right part of the distribution in Figs.~\ref{fig:sub_distribution1} (without considering the deprotonated cases) and~\ref{fig:sub_distribution2}. This makes sense because not all water molecules are equally strongly bound to the nanoparticle; indeed, Fig.~\ref{fig:distribution_h2o} presents also small BE values (around 15--30~kJ~mol$^{-1}$) and, accordingly, the multi H$_2$O adsorption will be the weighted average of all (strong and weak) binding sites.

\begin{table*}
    \centering
    \caption{Number of water molecules (nW) for each layer considered, and corresponding binding energies (BE, in~kJ~mol$^{-1}$). “A” and “N” stand for annealed and nucleated nanoparticles, respectively. The BE corresponds to the average BE of the outermost ML (e.g. of the 3rd ML for the grain with 3MLs). In the cases of proton ordered Ih (P-ice) and amorphous (Am-ice) ices the BE should be intended as the cohesive energy. Dipole moments in Debye. The size of the grain (in \AA) was calculated according to the radius of gyration.}
    \label{tab:appendix_BE_ML}
    \begin{tabular}{lccccccccc}
        \hline
Structure	& nAt$_{tot}$ &	$nW_{tot}$	&	$nW_{ML3}$	&	$nW_{ML2}$	&	$nW_{ML1}$	&	$nW_{OH}$	&	BE	&	Size	&	Dipole	\\
        \hline
A & 98 & -- & -- & -- & -- & -- & -- & 4.9 & 4.0\\
N & 98 & -- & -- & -- & -- & -- & -- & 5.1 & 11.0\\
\\
A - 1W	& 101 &	1	&		&		&		&	1	&	262.7	&	5.0	&	3.3	\\
N - 3W	& 108 &	3	&		&		&		&	3	&	179.4	&	5.3	&	11.0	\\
\\																			
A - 1ML	& 245 &	49	&		&		&	48	&	1	&	69.7	&	6.0		&	8.0	\\
N - 1ML	& 281 &	61	&		&		&	58	&	3	&	69.5	&	6.3		&	5.7	\\
\\																			
A - 2MLs & 527	&	143	&		&	94	&	48	&	1	&	40.4	&	7.5		&	9.7	\\
N - 2MLs & 557	&	153	&		&	92	&	58	&	3	&	36.3	&	7.9		&	16.8	\\
\\																			
A - 3MLs & 1235	&	379	&	236	&	94	&	48	&	1	&	39.1	&	9.9	&	24.5	\\
N - 3MLs & 1388	&	430	&	277	&	92	&	58	&	3	&	40.5	&	10.4	&	16.5	\\
\\																			
P-ice	& 216 &		&		&		&		&		&	51.5	&		&		\\
Am-ice	&	192	&		&		&		&		&	41.7	&		&		\\
        \hline
    \end{tabular}
\end{table*}

\subsection{Influence of water abundance on our snowlines}

\begin{figure*}
    \centering
    \includegraphics[width=0.8\linewidth]{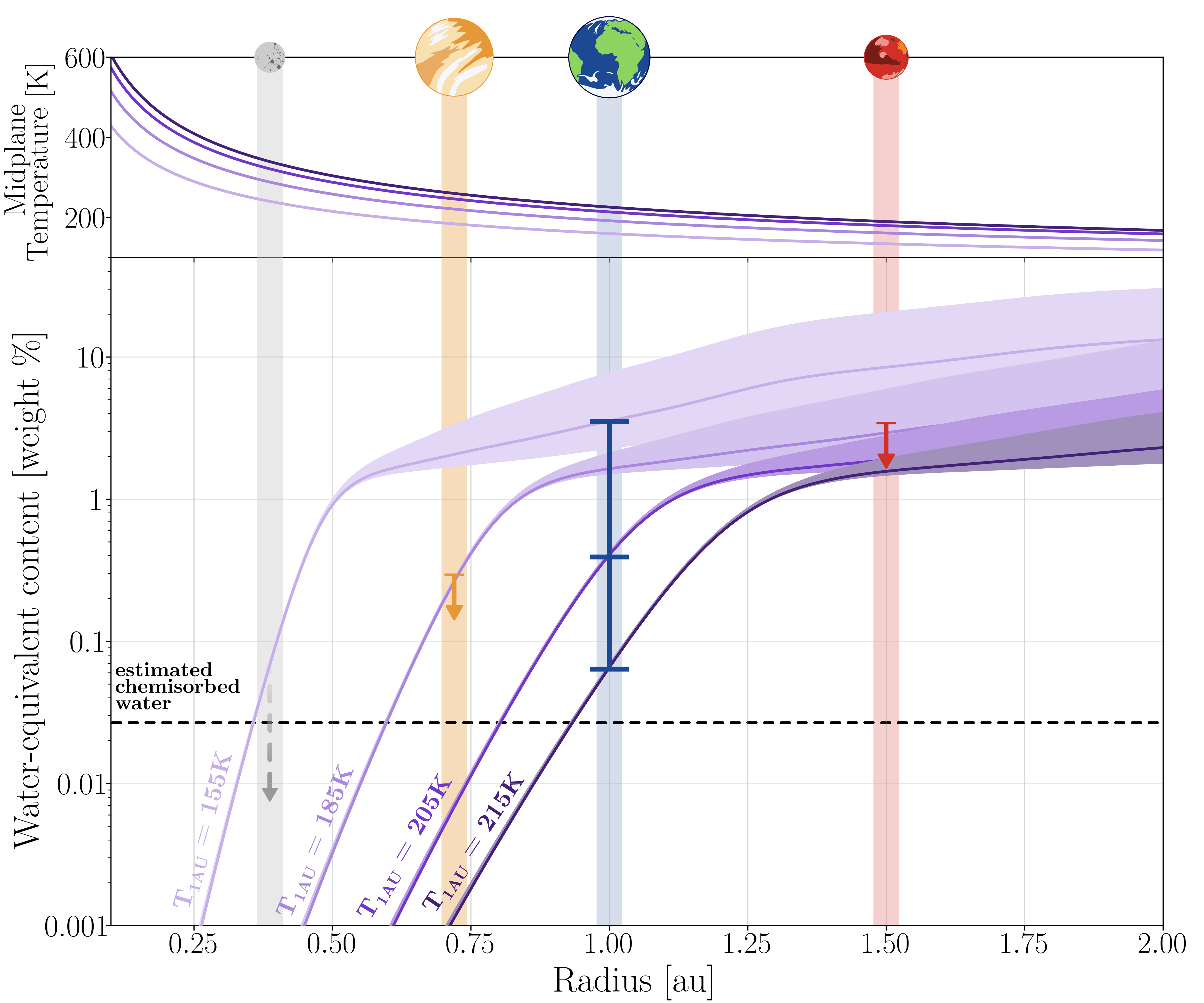}
    \caption{Influence of the chosen water abundance $A_{H_2O}$ on the snowlines from Figure \ref{fig:terrestrial_water}. The initial snowlines are kept as continuous purple lines. The shaded region around them correspond to a factor 2 higher and lower for $A_{H_2O}$}
    \label{fig:terrestrial_water_appendix}
\end{figure*}





\bsp	
\label{lastpage}
\end{document}